\documentclass[english,12pt,aps,pre,a4paper,preprintnumbers,floatfix,nofootinbib,showpacs,superscriptaddress,notitlepage]{revtex4-1}

\pdfoutput=1
\usepackage{subcaption}
\usepackage{graphicx}
\usepackage{rotate}
\usepackage[utf8]{inputenc}
\usepackage{bbold}
\usepackage{amsmath}
\usepackage{amssymb}
\usepackage{float}
\usepackage{comment}
\usepackage{soul}
\usepackage{lineno}
\usepackage[usenames,dvipsnames]{color}
\usepackage[colorinlistoftodos]{todonotes}
\usepackage[colorlinks=true, citecolor=blue, urlcolor=blue, linkcolor=blue]{hyperref}
\usepackage[normalem]{ulem}
\newcommand {\ignore}[1]{}
\usepackage{multirow}
\definecolor{darkred}{rgb}{0.6,0,0}

\newcommand{\rd}[1]{{\color{red}  #1}}
\newcommand{\bl}[1]{{\color{blue}  #1}}

\usepackage[english]{babel}
\usepackage{blindtext}
\definecolor{midnightblue}{RGB}{25,25,112}
\definecolor{brown}{rgb}{0.59, 0.29, 0.0}

\newcommand{\AddrIISERchem}{Department of Chemistry,
Indian Institute of Science Education and Research Bhopal 
Bhopal, Madhya Pradesh, 462066, India}

\newcommand{\AddrIISERphy}{Department of Physics,
Indian Institute of Science Education and Research Bhopal
Bhopal, Madhya Pradesh, 462066, India}

\begin{document}

\title{Non-Markovian escape under stochastic resetting} 

\author{Debasish Saha} 
\affiliation{\AddrIISERphy}
\author{Rati Sharma}\email{rati@iiserb.ac.in}
\affiliation{\AddrIISERchem}
\begin{abstract}
\vspace{0.5cm}
\noindent
Stochastic resetting is a powerful strategy known to optimize target-search processes at microscopic scales. While its effects on Markovian systems are well understood, its influence on memory-driven systems, such as in viscoelastic baths, has not been adequately investigated. In this work, we study the first-passage properties of escape for a harmonically trapped particle in a non-Markovian environment under stochastic resetting. We employ a complete renewal approach and find that the characteristic non-exponential heavy tail of the first passage time (FPT) distribution becomes exponential when resetting is introduced. We further find that optimal resetting is achievable at a lower reset rate when the dynamics are weakly correlated; however, for stronger correlations, the process needs to be reset more frequently. Therefore, resetting in memory-driven dynamics can be used as an effective control strategy to initiate faster escape, thereby regulating efficient transport mechanisms in complex chemical and biomolecular environments that follow non-Markovian dynamics.
\end{abstract}
\maketitle
\section{Introduction}
\label{sec:introduction}
\noindent
Chemical processes at the molecular level, such as enzyme kinetics, rapid bond formation and breaking and protein folding, play critical roles in various biological phenomena. The study of the kinetics of these chemical processes, therefore, provides valuable insights into the temporal evolution of the system, and how the system parameters govern these dynamics \cite{atkins2023, hanggi1986, hanggi1990}. Chemical kinetics, in its most general description, is described by a barrier-crossing formalism \cite{atkins2023, kramers1940, hanggi1986, hanggi1990}, in which reactant molecules undergo structural transformation from one state to another. The underlying mechanism is conceptualized as follows: a particle initially at a minimum (corresponding to an initial configuration) of a potential energy landscape gets activated by thermal fluctuations, and crosses the energy barrier to make a transition to another state (a different configuration). These energy landscapes are typically multi-dimensional spaces, consisting of a global minimum corresponding to the final stable configuration, and several local minima representing the metastable configurations. During a transition towards this stable configuration, a molecule passes through several such local minima by crossing over the intermediate energy barriers, being activated by thermal fluctuations, which are the dominant rules at this scale. 
\\~\\
To analytically tackle such problems, these barriers are often considered as a double-well potential, and the peak in the middle is considered as the barrier height. A particle is initially assumed to be in one of these wells (namely, the reactant well) and crosses over the barrier to reach the other well (product well). Regardless of how high the barrier is, there exists a non-zero probability of the particle escaping to the other side in a finite time \cite{atkins2023}. A lower barrier height allows faster transitions from reactants to products (high reaction rates), while a high barrier slows down the process (low reaction rates). A further simplified system is escape kinetics from a single potential well, in which particles diffuse stochastically inside the well, get absorbed by an absorbing wall, and permanently escape from the system. The escaping event is considered analogous to leaving the reactant well to form products. While diffusing, the particles gather kinetic energy from the surroundings, and when sufficiently energized, they escape the well in a finite interval of time. The average time taken by the particle to escape the well, \emph{i.e.}, the mean first-passage time (MFPT) or the mean escape time, serves as an indicator of the reaction time, and a factor characterizing the reaction rate. 
\\~\\
Classical models of barrier crossing or escape kinetics use Kramer's law to describe such processes under memoryless (Markovian) dynamics \cite{kramers1940, hanggi1986, cantisan2021, blumer2022, blumer2024}. However, many biochemical processes involving viscoelastic baths, such as rapid enzyme kinetics, protein folding, polymer translocation, and other multistep complex chemical processes, can often be explained via temporarily correlated dynamics, where memories of previous events affects future states \cite{mori1965, kubo1966, sung1996, lu1998, edman1999, flomenbom2005, min2005, min2006, english2006, chaudhury2006, goychuk2007, dubbeldam2007, chaudhury2007, wiggin2008, sharma2010, panja2010, chatterjee2010, vilk2024}. In fact, the rate constants, which are assumed to be constants in Markovian systems \cite{atkins2023}, become explicitly time-dependent when dealing with such viscoelastic environments \cite{chaudhury2006}. In such scenarios, non-Markovian models become necessary to explain these processes that exhibit correlated dynamics resulting from either complex multi-step reactions \cite{shahrezaei2008, sharma2012subdiffusion, roberts2015, batra2021, batra2022, vilk2024} or from dense environments that induce memory \cite{chaudhury2006, chatterjee2010, bhattacharyya2012, saha2024}. 
\\~\\
Escape kinetics have been extensively studied in both Markovian \cite{kramers1940, hanggi1986, cantisan2021, blumer2022, blumer2024} and non-Markovian \cite{chaudhury2006, goychuk2007, chatterjee2010} frameworks. In particular, a particle diffusing in viscoelastic baths is often described by the generalised Langevin equation (GLE) in which the friction coefficient is assumed to carry memory of past events, giving rise to dynamics that are inherently non-Markovian \cite{kou2004, min2005, min2006, goychuk2007, balakrishnan2008}. These memory effects, often modelled through time-dependent kernels, become crucial when dealing with complex chemical environments and anomalous diffusion \cite{min2005, min2006, goychuk2007, balakrishnan2008}. Earlier studies on bio-molecular interactions have shown power-law memory kernel to have very good agreement with experimental results, particularly for systems that exhibit anomalous diffusion \cite{min2005, min2006}. Later on, GLE with power-law memory kernel has been proven to be an effective model to describe complex kinetic phenomena, such as stochastic gene expression, immune-response, anomalous diffusion and protein folding, where the memory persists over time \cite{mason1995, min2005, chaudhury2006, chatterjee2010, chaudhury2007modulation, panja2010, sharma2010, chatterjee2011, bhattacharyya2012, thonnekottu2020, batra2021, batra2022}. 
\\~\\
Given the prevalence of non-Markovian behaviours, this study is particularly focused on accelerating the escape kinetics of a particle in a viscoelastic bath. In the past, several approaches have been utilized, such as fluctuating barriers \cite{doering1992, bier1993, reimann1998}, profile shaping \cite{chupeau2020}, and feedback \cite{coghi2025} to enable faster escape. In this work, we analyze the scope of a useful mechanism in non-equilibrium statistical mechanics, known as stochastic resetting \cite{evans2011, evans2011diffusion-optimal, evans2020}, within the context of non-Markovian dynamics. Stochastic resetting is a process in which the dynamics of a randomly diffusing particle is restarted from a given configuration at random intervals. The process is repeated several times until it finds the target or executes the given task \cite{evans2011, evans2011diffusion-optimal, pal2017, ray2020}. It has been observed that resetting can be used to optimize target search processes of stochastic dynamics by reducing search times (or FPTs) \cite{evans2011, evans2011diffusion-optimal, pal2015, pal2017, ray2020, ray2020space-dependent, tal2020, ginot2026}. The underlying mechanism can be explained as: starting from an initial configuration, the particle diffuses for some time $\tau$, and is reset to a position $x_{r}$ from where it continues to diffuse again. Time intervals at which resets happen are often taken from a Poissonian distribution, which means the reset intervals are completely random and independent of each other, and are also independent of the system's dynamics. This remarkable theory was proposed by Evans and Majumdar in 2011 \cite{evans2011}, which was experimentally verified in 2020 \cite{tal2020}. Apart from influencing the FPT behaviours, the process drives the system toward a non-equilibrium steady state in the long-time limit \cite{evans2011, gupta2022}. Because of these phenomena, stochastic resetting has been used in several research domains, including physics \cite{evans2020, gupta2022}, chemistry \cite{rotbart2015}, biology \cite{roldan2016, ramoso2020, jangid2026}, economics \cite{santra2022}, computer algorithms \cite{blumer2022, blumer2024} and many other applications where FPT is involved. 
\\~\\
Despite extensive studies, applications of resetting have largely been focused on Markovian systems. However, its suitability and potential applications in non-Markovian systems, in which dynamics are influenced by memory, are not yet well understood. Earlier, memory-dependent resetting protocols have been investigated, and different properties of diffusive dynamics emerge depending on the nature of the resetting protocol \cite{boyer2017}. More recent studies have been focused on understanding the dynamical effects of resetting in memory-driven systems \cite{tazbierski2025series, biswas2025, jolakoski2025, sandev2025, ginot2026}. In one such experimental study, it was shown that resetting can reduce the mean escape time of particles from a potential well in a non-Markovian environment \cite{ginot2026}. However, the role of memory, and whether resetting can be useful for optimizing escape kinetics, is still not well understood. Particularly, at what regime (weak or strong memory correlation) resetting can be beneficial has still remained a question. 
\\~\\
To investigate these, in this work, we utilize stochastic resetting in escape kinetics within a non-Markovian setup to examine whether it can accelerate the first-passage process, and whether it is possible to perform an optimal search. Previous attempts to understand the role of resetting in such escape mechanisms have focused on the Markovian systems \cite{ahmad2019, cantisan2021, singh2025}. In our study, the dynamics of the system are defined by the GLE, in which the viscous drag force has an explicit memory dependence. The GLE is then reduced to the Smoluchowski equation (or the generalized Fokker-Planck equation), representing the spatio-temporal evolution of probability density. Two additional terms appear in this equation -- the loss and gain of probabilities, due to reset \cite{evans2011, evans2011diffusion-optimal, evans2020}. The modified Smoluchowski equation is solved to find the probability density for position in order to have a macroscopic description of the system. Our results show that resetting can potentially improve escape mechanisms even though the bath tries to resist by applying memory-dependent viscous drag. The results demonstrate a transition from  memory-dominated to memoryless dynamics by incorporating stochastic resetting. This transition is reflected in the FPT distributions, characterized by a shift from non-exponential to exponential decay, indicating an effective memory loss. Memory usually slows down a process, and diminishing its effect by stochastic resetting makes the dynamics faster, and eventually accelerates the escape. However, the optimal escape mechanism strongly depends on the memory correlation strength, as stronger memory requires more frequent resetting, which can be energetically inefficient. Therefore, understanding the role of resetting in dense environments where dynamics are dominated by bath memory is of immense importance. There are potential applications, particularly in biological systems, which include rapid bond formation and bond-breaking, anomalous diffusion, transport through cell membranes, enzyme kinetics and drug delivery, among others, where conceptualizing the process in terms of resetting may be useful. This study can therefore provide insights into effective control mechanisms of reaction kinetics and transport in complex environments.
\\~\\
The manuscript is organized as follows. The mathematical formalism of the escape process and the underlying effects of memory and resetting on the dynamics are described in Sec. \ref{sec:formalism}. The first-passage properties of the system under resetting have been discussed in Sec. \ref{sec:first-passage-properties}. Key outcomes are discussed in Sec. \ref{sec:discussions}, and final conclusions are summarized in Sec. \ref{sec:conclusions}.
%
\section{Effect of resetting on system dynamics and memory}
\label{sec:formalism}
\noindent
Consider a particle of mass $m$, trapped in a potential well, exhibiting anomalous diffusion in a viscoelastic bath of temperature $T$. The dynamics of such a particle in the presence of temporally correlated noise and memory is governed by the one-dimensional generalized Langevin equation (GLE) \cite{mori1965, kubo1966}
\begin{equation}
\label{eq:gle}
    m\ddot{x}(t) = -\zeta \int_{0}^{t} dt^{\prime} K(t-t^{\prime}) \dot{x}(t^{\prime}) - \frac{dU(x)}{dx} + \theta(t),
\end{equation}
where, $x(t)$ is the particle's position at time $t$. The term on the left represents the force due to acceleration (or inertia). The first term on the right represents the viscous drag force generalized to include the non-Markovian regime by summing over the history for all times from the beginning. $\zeta$ here is the friction coefficient of the solvent, $K(t)$ is the memory kernel and $U(x)$ is the potential well under which the particle is trapped. $\theta(t)$ is the temporally correlated fractional Gaussian noise (fGn) with mean equal to zero, \emph{i.e.}, $\langle \theta (t) \rangle = 0$ \cite{mandelbrot1968, kou2004}. The fGn is related to the memory kernel $K(t)$ through the fluctuation-dissipation relation $\big\langle \theta(t) \theta(t^{\prime}) \big\rangle = \zeta k_{\text{B}}T K(|t-t^{\prime}|)$, where $K(|t-t^{\prime}|) = 2H(2H-1)|t-t^{\prime}|^{2H-2}$ \cite{kou2004}. Here $k_{\text{B}}$ is the Boltzmann constant, and $H$ is the Hurst index parameter ($1/2 \leq H < 1$) that defines the noise correlation strength, with higher values of $H$ corresponding to more strongly correlated dynamics. $H = 1/2$ corresponds to Gaussian white noise and represents the Markovian limit where the fluctuations become uncorrelated; whereas, for $H > 1/2$, the dynamics becomes non-Markovian, leading to subdiffusive dynamics \cite{desposito2009, thonnekottu2020}. 
\\~\\
In this study, we consider a harmonic potential well, \emph{i.e.}, $U(x) = \frac{1}{2}m\omega^{2}x^{2}$,  where $\omega$ is the well frequency. Considering overdamped limit, the above Eq. \eqref{eq:gle} can be modified as
\begin{equation}
\label{eq:gle-overdamped}
    m\omega^{2}x(t) = -\zeta \int_{0}^{t} dt^{\prime} K(t-t^{\prime}) \dot{x}(t^{\prime}) + \theta(t).
\end{equation}
The corresponding macroscopic properties of the system can be represented by the probability density, $P_{0}(x,t|x_{0})$, that the particle, starting from $x_{0}$ reaches a position $x$ at time $t$. The dynamics of this probability density is given by the Smoluchowski equation as \cite{okuyama1986, chaudhury2006}
\begin{equation}
\label{eq:fpe}
    \frac{\partial P_{0}(x,t|x_{0})}{\partial t} = \eta(t) \bigg[ \frac{\partial }{\partial x}x + \frac{k_{\text{B}}T}{m\omega^{2}} \frac{\partial^{2}}{\partial x^{2}} \bigg] P_{0}(x,t|x_{0}),
\end{equation}
with $P_{0}(x,0|x_{0}) = \delta(x-x_{0})$ as the initial condition. $\eta(t)$ is related to the time-dependent diffusion coefficient expressed as $\eta(t) = - \dot{\chi}(t) / \chi(t)$, in which $\chi(t) = E_{2-2H}\left[-\left(t / \tau_{0} \right)^{2-2H}\right]$, $\tau_{0} = \Big[ \frac{\zeta \Gamma(2H+1)}{m\omega^{2}} \Big]^{1/(2-2H)}$ is the relaxation time coefficient. $\Gamma(n)$ is the Gamma function, and $E_{\alpha}(-z) = \sum_{k=0} ^{\infty} (-1)^{k} \frac{z^{k}}{\Gamma(\alpha k + 1)}$ is the Mittag-Leffler function \cite{haubold2011}. 
\\~\\
The dynamics is intermittently restarted from the initial configuration at a constant rate $r$, and at a random time interval $\tau$ drawn from a Poissonian distribution $re^{-r\tau}$ \cite{evans2011, evans2011diffusion-optimal}. In the Markovian case, only position resetting is sufficient, as the system does not have any memory. In contrast, for a non-Markovian system, position resetting alone is not sufficient, as the residual drag force (the first term on the right side of Eq. \eqref{eq:gle-overdamped}) will push the particle back to the pre-resetting position \cite{ginot2026}. Therefore, at every resetting event, the bath has to be relaxed as well, so that the previous memory is completely erased, and a new process (completely independent of the previous one) begins \cite{tazbierski2025series, biswas2025, ginot2026}. To incorporate that, we consider a complete renewal technique, where the position of the particle as well as the memory kernel is reset to the initial configurations. 
Then the Smoluchowski equation, under stochastic resetting, can be modified as
\begin{equation}
\label{eq:fpe-with-resetting}
    \frac{\partial P_{r}(x,t|x_{0})}{\partial t} = \eta(t) \bigg[ \frac{\partial }{\partial x}x + \frac{k_{\text{B}}T}{m\omega^{2}} \frac{\partial^{2}}{\partial x^{2}} \bigg] P_{r}(x,t|x_{0}) - rP_{r}(x,t|x_{0}) + r\delta(x-x_{r}),
\end{equation}
where $P_{r}$ represents the probability density in the presence of resetting. The second term on the right-hand side shows the loss in probability from an arbitrary position $x$ due to reset, while the last term represents the gain in probability at the resetting position $x_{r}$. The system develops a non-zero current around the reset position, giving rise to a non-equilibrium steady state at large times. It should be noted that at every reset event, the history of previous memory is completely erased, and fluctuations become uncorrelated from one diffusive phase to another. It is also assumed that during a reset, the bath relaxes instantaneously. But, in an experimental setup, a real non-Markovian system should be given a sufficient amount of time to relax the bath completely \cite{ginot2026}. Between two successive resets, the dynamics remain non-Markovian, whereas the resetting events follow Markovian properties as they are assumed to be independent of previous resets. 
\\~\\
Throughout the study, we have considered the following values of the parameters for evaluation. The mass of the particle ($m$), friction coefficient ($\zeta$), trap frequency ($\omega$), Boltzmann constant ($k_{\text{B}}$), and bath temperature ($T$) are fixed to unity (in arbitrary units). The Hurst index ($H$) is varied between 0.5 and 0.95. The process starts from the initial position $x_{0}=0$, \emph{i.e.}, from the bottom of the well. Dynamics under resetting are checked for different reset rates, which we vary between 0 and 10. Different reset positions are taken in different situations. For example, in the study of position distributions, we choose $x_{r} = 1$, whereas for evaluating mean-squared displacement, correlation function and first-passage properties, we set $x_{r} = 0$. These values are kept consistent for both analytical and numerical evaluations. Quantities that are varied in different cases are mentioned in every figure caption. 

\subsection{Position probability density}
\label{sec:position-probability-density}
\begin{figure}
    \centering
    \includegraphics[width=\linewidth]{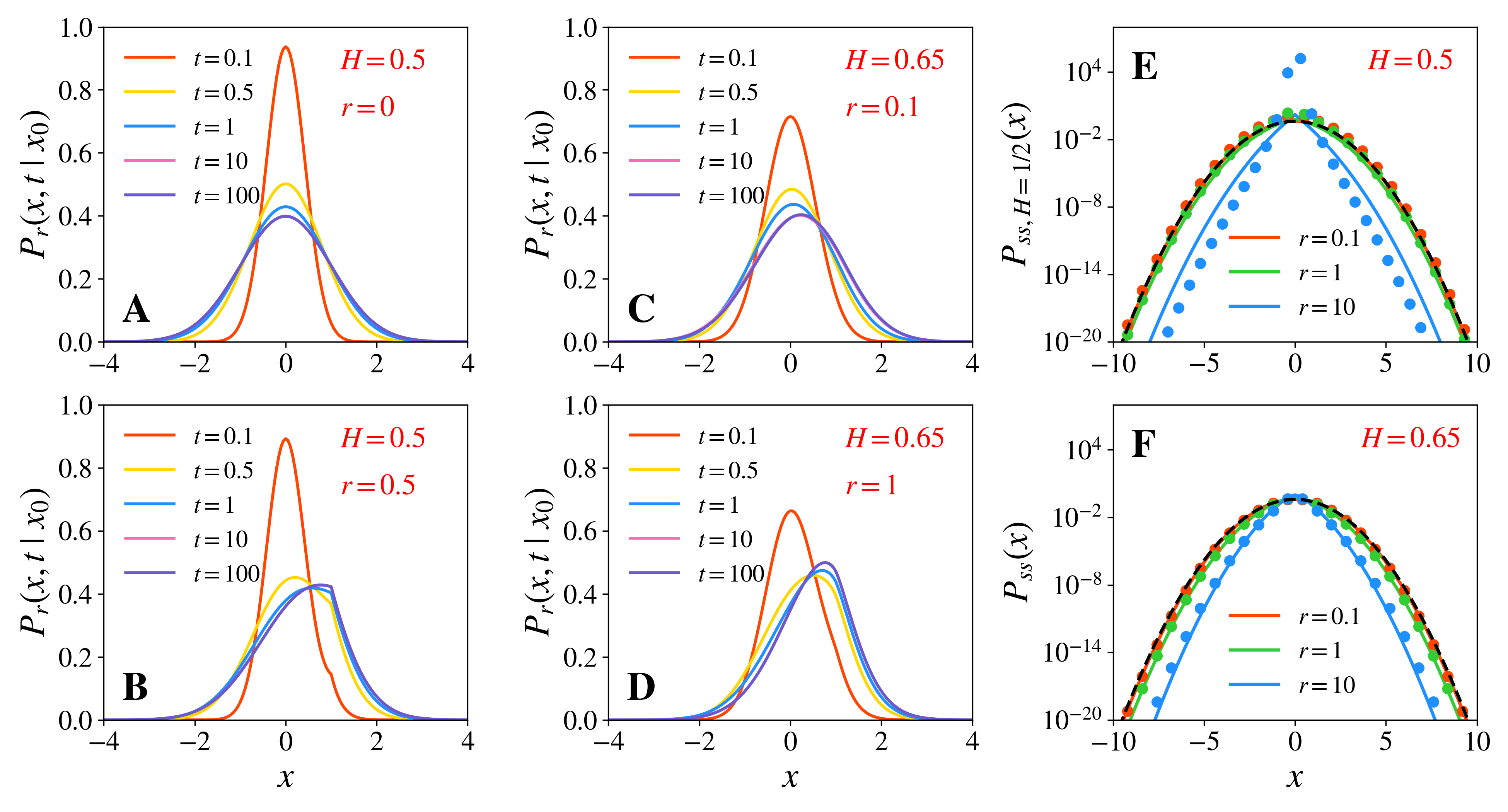}
    \caption{Probability density for a particle in a viscoelastic bath governed by GLE and under a harmonic trap, while being subjected to stochastic resetting. Distributions are shown at different Hurst indices ($H$) and reset rates ($r$) to elucidate their effects on the dynamics, and at different times to illustrate the emergence of the stationary state. (A) and (B) show the probability density with respect to position for the Markovian case ($H=0.5$) in the absence ($r=0$, in Fig. A) and presence ($r=0.5$, in Fig. B) of resetting. (C) and (D) show the same for the non-Markovian dynamics ($H=0.65$) and in the presence of low ($r=0.1$, Fig. C) and relatively high ($r=1$, Fig. D) reset rates. The reset position was chosen to be $x_{r}=1$ in all the cases. (E) Steady-state probability density at the Markovian limit for different reset rates. The solid lines represent the exact probability density obtained from the renewal equation, Eq. \eqref{eq:stationary-state}, and the dots represent the approximated NESS density, as calculated from Eq. \eqref{eq:P-NESS-white-final}. The black dashed line is the Gaussian $\textrm{e}^{-x^{2}/2}/\sqrt{2\pi}$, shown for comparison. (F) Steady-state probability density for the non-Markovian case ($H = 0.65$), with lines and dots having the same meaning as in Fig. E. Lines are obtained from Eq. \eqref{eq:stationary-state} and dots are obtained from Eq. \eqref{eq:P-NESS-Final}. The reset position is $x_{r} = 0$ for both Figs. E and F.}
    \label{fig:prob-density-position}
\end{figure}

\noindent
Let us first understand the effects of resetting on the dynamics of the system. It has been observed that there exists a non-equilibrium steady state when the dynamics are restarted at Poissonian intervals \cite{evans2011, evans2011diffusion-optimal, evans2020, gupta2022}. To investigate this, we first solve Eq. \eqref{eq:fpe-with-resetting} to obtain $P_{r}(x,t|x_{0})$, in the presence of resetting at a constant rate. The solution can be obtained following the last renewal approach \cite{evans2011diffusion-optimal, evans2020}, which is given by 
\begin{equation}
\label{eq:renewal-equation}
    P_{r}(x,t|x_{0}) = e^{-rt} G(x,t|x_{0},0) + r\int_{0}^{t} d\tau ~e^{-r\tau} G(x,\tau|x_{r}).
\end{equation}
The first term on the right-hand side corresponds to the trajectories that experience no resetting in time $t$, with no-reset probability $e^{-rt}$. The second term represents the trajectories, in which the last resetting occurs at time $t -\tau$, followed by diffusion for the rest of the interval of duration $\tau$, starting from the resetting position $x_{r}$, and integrating over all $d\tau$. Here $G(x,t|x_{0},0)$ is the propagator, defining the probability of transition from $x_{0}$ at time $t=0$ to $x$ at a later time $t$. It should be noted that the renewal equation holds under complete renewal of the dynamics, \emph{i.e.}, resetting the position as well as the bath memory. The propagator, $G(x,t|x_{0},0)$ satisfies the equation
\begin{equation}
\label{eq:green-function-equation}
    \bigg( \frac{\partial}{\partial t} - \mathcal{L} \bigg) G(x,t-t^{\prime}|x^{\prime}) = \delta(x-x^{\prime}) \delta(t-t^{\prime})
\end{equation}
where $\mathcal{L} \equiv \eta(t) \Big[ \frac{\partial }{\partial x}x + \frac{k_{B}T}{m\omega^{2}} \frac{\partial^{2}}{\partial x^{2}} \Big]$. The solution of this Eq. \eqref{eq:green-function-equation} is \cite{batra2021}
\begin{equation}
\label{eq:greens-function}
    G(x,t|x_{0},0) = \sqrt{\frac{m\omega^{2}}{2\pi k_{\text{B}}T\big(1-\chi^{2}(t)\big)}} \exp{\Bigg[ -\frac{m\omega^{2}\big(x-x_{0}\chi(t)\big)^{2}}{2 k_{\text{B}}T\big(1-\chi^{2}(t)\big)} \Bigg]}.
\end{equation}
Next, we solve Eq. \eqref{eq:renewal-equation} to understand the underlying dynamics and the impact of resetting. However, getting a closed-form expression of $P_{r}(x,t|x_{0})$ is non-trivial because of the complicated expression of the propagator. Therefore, we solve Eq. \eqref{eq:renewal-equation} numerically in Mathematica \cite{Mathematica}, and plot $P_{r}(x,t|x_{0})$ for different values of the Hurst index ($H$) and reset rate ($r$), which are shown in Fig. \ref{fig:prob-density-position}. Probability densities are shown at different times to illustrate how the system evolves and eventually reaches a steady state (SS) at long times under different reset conditions. In the Markovian limit ($H=0.5$), the system reaches the SS at large times, as shown in Figs. \ref{fig:prob-density-position}\bl{A} and \ref{fig:prob-density-position}\bl{B}. In the absence of resetting, distributions spread symmetrically around the potential minimum at short times, and converge to the SS, following the standard Ornstein-Uhlenbeck process \cite{balakrishnan2008}, as shown in Fig. \ref{fig:prob-density-position}\bl{A}. On the other hand, when the dynamics undergo resetting at a position away from the minimum ($x_{r}=1$), a net drift in $P_{r}(x,t|x_{0})$ is observed towards the reset position, and reaches an SS, as shown in Fig. \ref{fig:prob-density-position}\bl{B}. 
\\~\\
In Figs. \ref{fig:prob-density-position}\bl{C} and \ref{fig:prob-density-position}\bl{D}, position distributions under resetting for the non-Markovian case ($H = 0.65$) are shown. Similar to the Markovian process, the system again reaches an SS at long times.
In the figures, we have compared the effects of resetting on the distribution at different reset rates. For a small reset rate ($r = 0.1$), the peak is formed in a region between the well minimum and reset position ($x_{r} = 1$ in this case). The resetting process moves the particle towards $x_{r}$, whereas the harmonic trap tries to bring it near its minimum, and the competition between these two determines the resulting fate of the dynamics. Resetting at a low rate indicates that the dynamics are restarted less frequently, and the system gets sufficient time to drift away from the reset position. Similarly, when the dynamics are reset at a relatively higher reset rate, the distribution peak is formed near the reset position, as the particle has less time to diffuse away. One subtle difference is that, unlike the Markovian case, there is no sharp peak in the position distribution of the non-Markov process (as shown in Figs. \ref{fig:prob-density-position}\bl{C} and \ref{fig:prob-density-position}\bl{D}). A possible reason is the inherent temporal correlation in the dynamics as described in Eq. \eqref{eq:gle-overdamped}. According to this, the dynamics of a non-Markovian process are obtained by cumulating the viscous force over a given time interval, and because of that, the sharp characteristics that are supposed to appear near the reset position are smoothed out. 
\\~\\
The SS itself can be obtained by taking the limit $t \to \infty$ in Eq. \eqref{eq:renewal-equation}, which results in
\begin{equation}
\label{eq:stationary-state}
    P_{\text{ss}}(x|x_{0}) = r\int_{0}^{\infty} d\tau ~e^{-r\tau} G(x,\tau|x_{r}).
\end{equation}
Although the above relation, Eq. \eqref{eq:stationary-state}, cannot be solved exactly, one can deduce the approximate expression for the non-equilibrium steady state (NESS) via the saddle point approximation method \cite{majumdar2015dynamical, evans2020}. 
For the case of $H=1/2$, corresponding to white noise, $\chi(t) = \textrm{exp}(-a_1 t)$, which leads to $[1-\chi^2(t)]^{-1} \sim 1+e^{-2a_1 t}$ for long times after binomial expansion, with $a_1 = m\omega^2/\zeta$. Substituting this form of $\chi(t)$ in Eq. \eqref{eq:greens-function}, gives
\begin{equation}
    G(x,t|0,0) = \sqrt{\frac{m\omega^2}{2\pi k_\mathrm{B}T}}\left(1-e^{-2a_1 t}\right)^{-1/2} ~\mathrm{exp}\left[\frac{-m\omega^2}{2k_\mathrm{B}T}x^2(1+e^{-2a_1 t}) \right]
    \label{eq:greens-white}
\end{equation}
Considering $x_r = 0$, Eq. \eqref{eq:stationary-state} can be written as
\begin{equation}
    P_{ss,H=1/2} \sim \int_0^\infty d\tau \textrm{ exp}[-\phi_{H=1/2}(\tau,x)]
    \label{eq:P-NESS-white}
\end{equation}
where, $\phi_{H=1/2}(\tau,x) = r\tau + Ax^2(1+e^{-2a_1\tau})$, with $A = \frac{m\omega^2}{2k_\mathrm{B}T}$. Now, for large $|x|$ and large $t$, one can carry out the saddle point approximation. The integral in Eq. \eqref{eq:P-NESS-white} will have the largest contribution from $e^{-\phi_{H=1/2}(\tau, x)}$ when $\phi_{H=1/2}(\tau, x)$ is minimum, which is found to be at $\tau^*(x) = \frac{1}{2a_1}\textrm{ln}\left(\frac{2Ax^2a_1}{r}\right)$. This then leads to the approximate NESS, given by
\begin{align}
    P_{ss,H=1/2} &\sim \textrm{exp}[-\phi_{H=1/2}(\tau^*,x)] \nonumber \\
    &\sim \textrm{exp}(-Ax^2)|x|^{-r/a_1}
    \label{eq:P-NESS-white-final}
\end{align}
The NESS for the $H=1/2$ case is therefore a Gaussian multiplied by an algebraic function to the decreasing power of $r/a_1$. An analysis of this equation suggests that the Gaussian term dominates when $r<<a_1$. The resetting NESS therefore starts to have an effect only when $r\geq a_1$. This indicates that a temporal regime $a_1^{-1}< t <r^{-1}$ exists where the particle is in equilibrium with resetting-induced NESS established only when $t \geq r^{-1}$. The resetting-induced NESS for $H=1/2$, as given in Eq. \eqref{eq:P-NESS-white-final}, is compared to the numerically computed steady state from Eq. \eqref{eq:stationary-state} in Fig. \ref{fig:prob-density-position}\bl{E}. For smaller values of $r$, the steady state is nearly Gaussian. The approximate NESS, given by Eq. \eqref{eq:P-NESS-white-final} matches well with Eq. \eqref{eq:stationary-state}. The presence of the singularity in the NESS expression is reflected in the divergence around $x=0$, as depicted in the figure. Therefore, the NESS is only valid at large $|x|$ and at large $t$. 
\\~\\
Now, we analyze the case of the fGn, \textit{i.e.}, when $H \neq 1/2$. At long times, $\chi(t) \sim a_2 t^{-b}$ \cite{erdelyi1953higher, chatterjee2010, chatterjee2011}, which leads to $[1-\chi^2(t)]^{-1} \sim 1+a_2^2t^{-2b}+\mathcal{O}(t^{-4b})$ by binomial expansion. Here, $a_2 = \frac{\Gamma(2H+1)}{a_1\Gamma(2H-1)}$ and $b=2-2H$. This form of $\chi(t)$ then renders the propagator, $G(x,t|0,0)$ from Eq. \eqref{eq:greens-function} to
\begin{equation}
    G(x,t|0,0) = \sqrt{\frac{m\omega^2}{2\pi k_\mathrm{B}T}}(1-a_2^2t^{-2b})^{-1/2} \mathrm{exp}\left[\frac{-m\omega^2}{2k_\mathrm{B}T}x^2(1+a_2^2t^{-2b}) \right]
    \label{eq:greens-approximate}
\end{equation}
Now, for large $|x|$ and large $t$, we can obtain the asymptotic tail of the steady state distribution (Eq. \eqref{eq:stationary-state}) via the saddle point approximation. Considering $x_r=0$ for simplicity, Eq. \eqref{eq:stationary-state} reduces to
\begin{equation}
    P_{ss}(x) \sim \int_0^\infty d\tau e^{-\phi(\tau,x)}
    \label{eq:P-NESS}
\end{equation}
where, $\phi(\tau,x) = r\tau + Ax^2(1+a_2^2\tau^{-2b})$. As before, the integral in Eq. \eqref{eq:P-NESS} will have the largest contribution from $e^{-\phi(\tau,x)}$ when $\phi(\tau,x)$ is the minimum. Therefore, setting $d\phi/d\tau = 0$ gives the minimum at $\tau^*(x) = (2bAa_2^2x^2/r)^\frac{1}{2b+1}$. This then leads to the approximate NESS
\begin{align}
    P_{ss}(x) &\sim \mathrm{exp}[-\phi(\tau^*,x)] \nonumber \\
    &\sim \textrm{exp}(-Ax^2)\textrm{exp}(-a_3|x|^{\frac{1}{b+1/2}})
    \label{eq:P-NESS-Final}
\end{align}
where, $a_3 = \frac{r(2b+1)}{2b}\left(\frac{2bAa_2^2}{r} \right)^\frac{1}{2b+1}$. The NESS, is therefore, a Gaussian multiplied by a stretched exponential function. Similar to the case for $H=1/2$, there exists a temporal regime $a_2^{1/b} < t < r^{-1}$ where the system is in equilibrium, with the resetting NESS being established only when $t \geq r^{-1}$. Again, as shown in Fig. \ref{fig:prob-density-position}\bl{F}, the approximate analytical expression for the NESS distribution, Eq. \eqref{eq:P-NESS-Final}, fits well with the complete expression, Eq. \eqref{eq:stationary-state}, computed numerically. It can be seen that the NESS deviates from Gaussian for large values of $r$, as the effect of resetting becomes more prominent. 
\\~\\
The change in the probability density and the subsequent formation of SS at the reset position is one of the remarkable characteristics of stochastic resetting, which was observed in several other systems \cite{evans2011, evans2011diffusion-optimal, roldan2017, gupta2022, pal2015, tal2020, evans2020}. As we reset the dynamics, a source of probability is generated at the reset position, and the distribution gradually shifts towards the reset position. At the trajectory level, the particles diffuse in the neighbourhood of the reset position, as it gets reset before it can go large distances. As a result, a net probability current is generated, leading to the non-equilibrium SS. 
\begin{figure}
    \centering
    \includegraphics[width=0.8\linewidth]{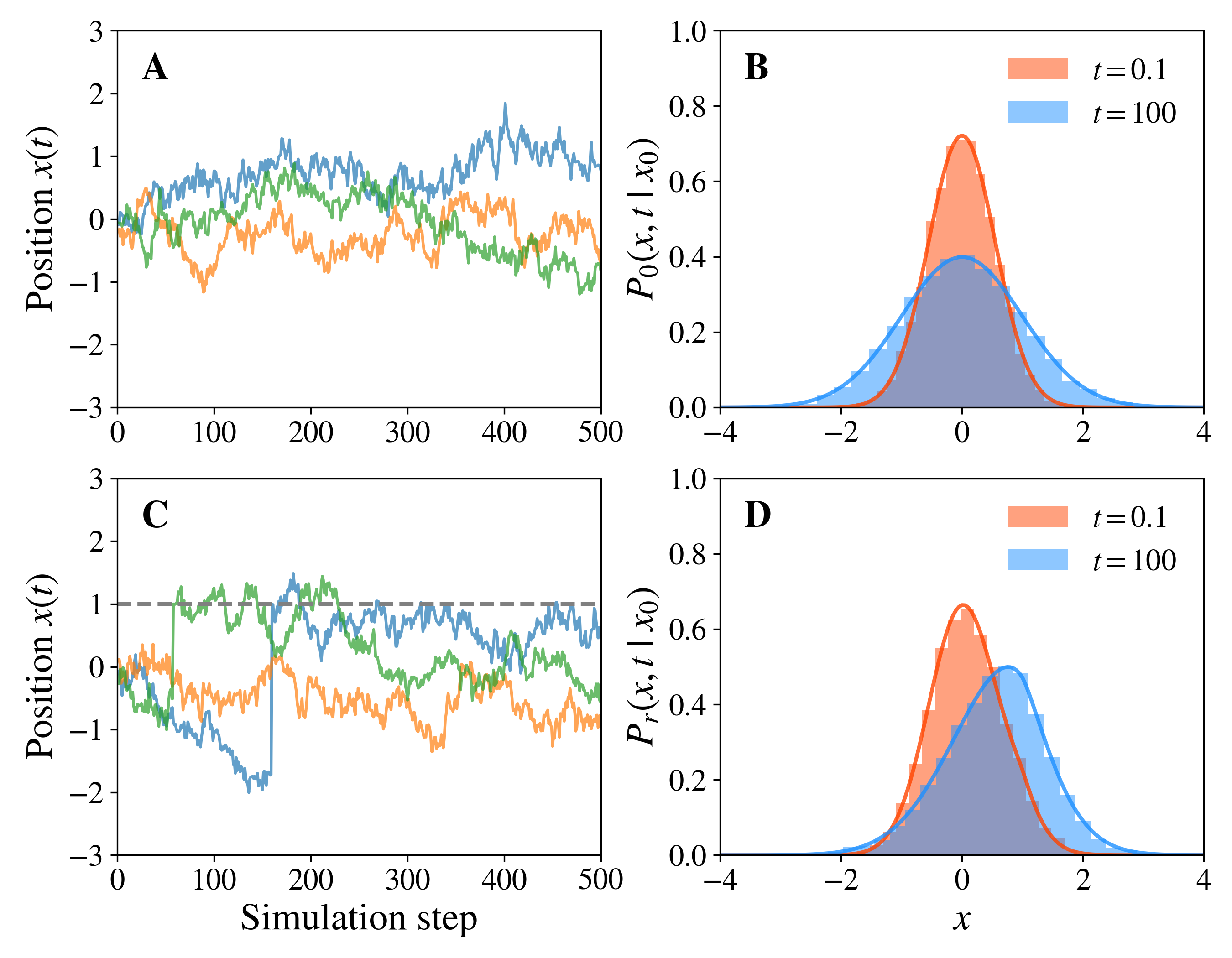}
    \caption{Trajectories and corresponding distributions without and with resetting. (A) Sample trajectories of particles following the dynamics described by Eq. \eqref{eq:gle-overdamped} and \eqref{eq:final-numerical-solution} without resetting ($r=0$). (B) Comparison between the analytical distributions (solid lines) and simulated histograms for $r = 0$. (C) Sample trajectories showing diffusion with stochastic resetting ($r=1$), where the jumps on the horizontal dashed line at $x_{r}=1$ indicate the resetting events. The follow-up diffusive phase starts from $x_{r}$ after every reset. (D) Comparison between the analytical and numerical results of the probability distributions in the presence of resetting ($r=1$). Trajectories and histograms are obtained by simulating $10^{4}$ trajectories with $\Delta t = 10^{-3}$, keeping $H=0.65$ and $x_{r}=1$.}
    \label{fig:position-dist-simulation}
\end{figure}
\\~\\
To validate analytical results, we performed numerical simulations of the process, following Eq. \eqref{eq:final-numerical-solution}, a discretized form of Eq. (\ref{eq:gle-overdamped}), and testing how stochastic resetting influences the dynamics. The discretization method and simulation techniques are described in Appendix \ref{sec:numerical-method} and \ref{app:simulation-steps}, respectively. Sample trajectories and the corresponding position distributions are shown in Fig. \ref{fig:position-dist-simulation}. The distributions are plotted at different times, which are generated from $10^{4}$ trajectories, with a time interval $\Delta t = 10^{-3}$. Other parameters were kept fixed at the following values: $H = 0.65$ and $x_{r} = 1$. Figs. \ref{fig:position-dist-simulation}\bl{A} and \ref{fig:position-dist-simulation}\bl{B} show the trajectories and histograms, respectively, corresponding to the no-resetting ($r = 0$) case. The simulated distributions show good agreement with analytical results obtained from the renewal equation provided in Eq. \eqref{eq:renewal-equation}. Fig. \ref{fig:position-dist-simulation}\bl{C} illustrates the effect of resetting, where every sudden jump to the reset position (black dashed line) represents the resetting events of individual trajectories. After every reset, the follow-up diffusion process starts from the reset position. The corresponding distributions for the position of the particles are shown in Fig. \ref{fig:position-dist-simulation}\bl{D}. It can be seen how the distribution shifts with time as we reset the dynamics. This distribution also shows good agreement with the analytically obtained distributions at both time instants.

\subsection{Mean-squared displacement}
\label{sec:msd}
\begin{figure}
    \centering
    \includegraphics[width=0.9\linewidth]{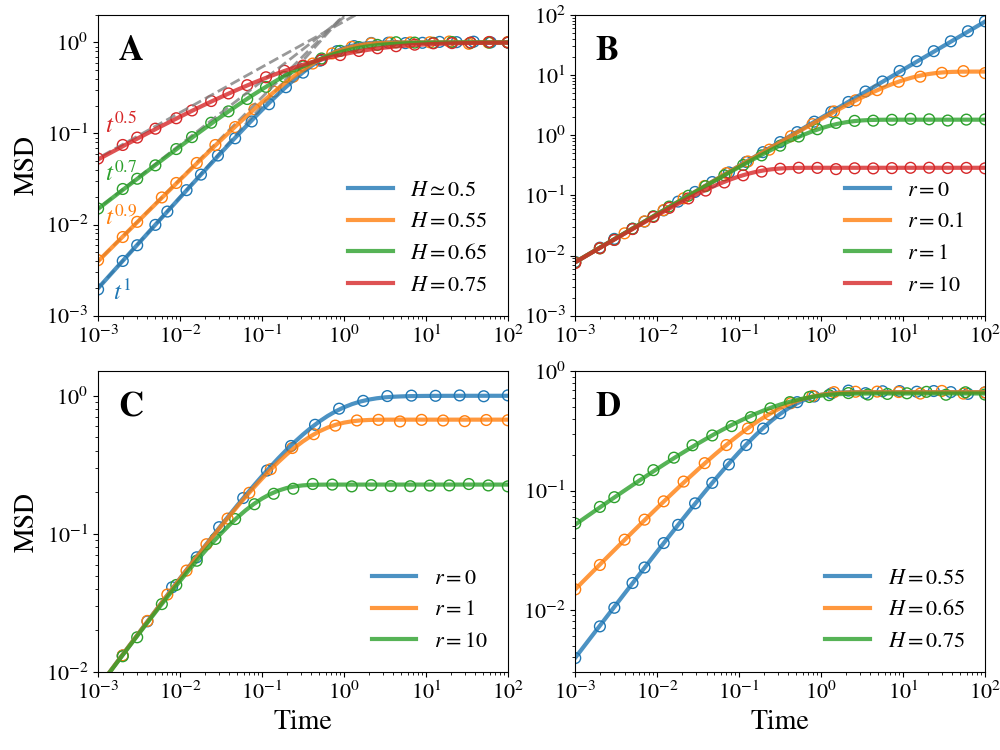}
    \caption{Mean-squared displacement of a particle diffusing in a viscoelastic bath and harmonic trap, under resetting. (A) Mean-squared displacement without resetting. Different colours correspond to different Hurst indices ($H$). The grey dashed lines represent the cases when no harmonic trap was considered ($\omega = 0$), showing characteristics of free subdiffusion. Otherwise, for coloured plots, $\omega$ is considered to be unity. The slopes at a small time limit (and for free subdiffusion) are $\sigma^{2}(t) \sim t^{2-2H}$. (B) MSD at different values of reset rate in the absence of a harmonic trap ($\omega = 0$). (C) MSD at different reset rates and in the presence of a harmonic trap ($\omega = 1$). (D) MSD of the particle under the trap at a fixed reset rate ($r = 1$), but at different Hurst indices. The empty circles represent the MSD estimated from simulated trajectories. Hurst index is 0.6 for both Figs. B and C. Simulated MSDs are estimated from $10^{4}$ trajectories with $\Delta t = 10^{-3}$.} 
    \label{fig:msd}
\end{figure}

\noindent
The emergence of a stationary state is evident from the position distributions as discussed in Sec. \ref{sec:position-probability-density}. Two distinct natures of the stationary states are observed -- one is an equilibrium stationary state, which appears due to the stable potential well, and can be observed even when no resetting is considered. The other is a non-equilibrium stationary state (NESS) arising from the probability current induced by resetting. To understand these different regimes and the associated timescales of the dynamics, we study the mean-squared displacement (MSD) of the process. Before calculating the MSD under resetting, we first need to evaluate the MSD of the original process (without any resetting).
\\~\\
As previously reported \cite{desposito2009, biswas2025}, the MSD of a particle in a viscoelastic bath and trapped under a harmonic potential can be expressed as (from Appendix \ref{app:msd_derivation})
\begin{equation}
\label{eq:msd_no_reset}
    \sigma^{2}(t) = \frac{k_{\text{B}}T}{\zeta} \Big[ 2I(t) - \omega'^{2}I^{2}(t) \Big],
\end{equation}
where, $\omega'^{2} = m\omega^{2}/\zeta$, and $I(t)$ is a relaxation function defined as 
\begin{equation}
\label{eq:relaxation_function_def}
    I(t) = \mathcal{L}^{-1}\big[ \hat{I}(s) \big] = \mathcal{L}^{-1}\bigg[ \frac{s^{-1}}{s \hat{K}(s) + \omega'^{2}} \bigg],
\end{equation}
where, $\mathcal{L}^{-1}\big[ \cdot \big]$ represents the inverse Laplace transform. Here, $\hat{K}(s)$ is the Laplace transform of the memory kernel $K(t)$. In our case, we have considered a power-law memory kernel, represented as
\begin{equation}
\label{eq:memory_kernel}
    K(t) = a t^{-b}, 
\end{equation}
where $a = 2H(2H-1)$ and $b = 2-2H$. Taking the Laplace transform of the kernel, which is $\hat{K}(s) = a\Gamma(1-b)/s^{1-b}$, and using this in Eq. \eqref{eq:relaxation_function_def}, we obtain 
\begin{equation}
\label{eq:relaxation_function}
    \hat{I}(s) = \frac{s^{-1}}{A_{1}s^{b} + \omega'^{2}}, 
\end{equation}
with $A_{1} = a\Gamma(1-b) = 2H(2H-1) \Gamma(2H-1) = \Gamma(2H+1)$. 
\\~\\
Now, to take the inverse Laplace transform of $\hat{I}(s)$, we use the property of Mittag-Leffler function with parameters $\alpha, \beta > 0$, \emph{i.e.}, $\mathcal{L}[t^{\beta-1} E_{\alpha, \beta}(-\lambda t^{\alpha})] = \frac{s^{\alpha-\beta}}{s^{\alpha} + \lambda}$, where $E_{\alpha, \beta}(z)$ is defined as $E_{\alpha, \beta} = \sum_{k=0}^{\infty} z^{k} / \Gamma(\alpha k + \beta)$. The Laplace inversion leads to
\begin{equation}
    I(t) = \frac{t^{b}}{A_{1}} E_{b, b+1}\bigg( -\frac{\omega'^{2}}{A_{1}} t^{b} \bigg).
\end{equation}
We use the Mittag-Leffler identity $z E_{b, b+1} (z) = E_{b}(z) - 1$, substitute $z = -\lambda t^{b}$ to obtain 
\begin{equation}
    t^{b} E_{b, b+1} (-\lambda t^{b}) = \frac{1}{\lambda} \left[ 1 - E_{b}(-\lambda t^{b}) \right] ,
\end{equation}
where, $\lambda = \omega'^{2}/A_{1}$. Making use of this, we can write the relaxation function as a function of a single parameter Mittag-Leffler function
\begin{equation}
    I(t) = \frac{1}{\omega'^{2}} \left[ 1 - E_{b} \left(-\frac{\omega'^{2}}{A_{1}} t^{b} \right) \right].
\end{equation}
Putting this back in Eq. \eqref{eq:msd_no_reset} and using the value of $\omega'^{2}$, we get the expression of MSD with a power-law memory kernel
\begin{equation}
\label{eq:msd_no_reset_omega_neq_zero}
\begin{split}
    \sigma^{2}(t) &= \frac{k_{\text{B}}T}{m \omega^{2}} \bigg[ 1 - \left\{ E_{b} \left(-\frac{m\omega^{2}}{\zeta A_{1}} t^{b} \right) \right\}^{2}  \bigg] \\
    &= \frac{k_{\text{B}}T}{m \omega^{2}} \left[ 1 - \chi^{2}(t) \right]
\end{split}
\end{equation}
\\
In the case of free diffusion, we use $\omega = 0$ in the above equation, but that would cause the MSD to diverge. To avoid that, we set $\omega'$ to zero directly in Eq. \eqref{eq:relaxation_function}, and then take inverse Laplace transform to get the relaxation function for free diffusion, which is 
\begin{equation}
    I(t)_{\omega=0} = \frac{t^{b}}{A_{1} \Gamma(1+b)}.
\end{equation}
Finally, we get the MSD for free diffusion as, 
\begin{equation}
\label{eq:msd_no_reset_omega_eq_zero}
    \sigma^{2}(t) = \frac{2 k_{\text{B}}T}{\zeta A_{1} \Gamma(1+b)} ~t^{b},
\end{equation}
which shows $\sigma^{2}(t) \sim t^{2-2H}$, by putting the value of $b$. In the Markovian limit ($H=1/2$), the MSD varies as $\sigma^{2} \sim t$ \cite{balakrishnan2008}. In the non-Markovian limit ($1/2 < H < 1$), it varies as $\sigma^{2} \sim t^{\alpha}$ with $0< \alpha < 1$, which is a characteristic signature of anomalous diffusion, typically subdiffusion \cite{dubbeldam2007, desposito2009, panja2010, sharma2012subdiffusion}. 
\\~\\
The corresponding MSDs at different noise correlations are shown in Fig. \ref{fig:msd}\bl{A}. For free diffusion (in a viscoelastic bath), MSDs are linear in the log-log scale, as shown by the grey dashed lines, but with varying slopes ($\sim t^{\alpha}$) depending on the correlation strengths, as described above. Irrespective of that, all the dashed lines continue growing linearly, indicating that the system (without a trap) remains in the diffusive phase at every time scale. Interesting phenomena can be seen when we place the particle in a trap, which attempts to confine its dynamics and restrict its motion in the vicinity of the minimum potential. As a result, a steady state emerges in the long-time limit, whereas at short times it exhibits subdiffusion, as the slopes coincide with the grey dashed lines. But for a given trap stiffness ($\omega=1$), the time to reach the stationary state strongly depends on the noise correlation, defined by $H$, as shown in Figs. \ref{fig:msd}\bl{A} (coloured plots) and \ref{fig:msd_saturation}\bl{B}. In the Markovian limit, where fluctuations are uncorrelated, the MSD saturates very rapidly. But with increasing correlation, the system takes longer to get to the saturation regime. 

\subsubsection{MSD under resetting}
\label{subsec:msd_under_resetting}
\begin{figure}[htbp]
    \centering
    \includegraphics[width=\linewidth]{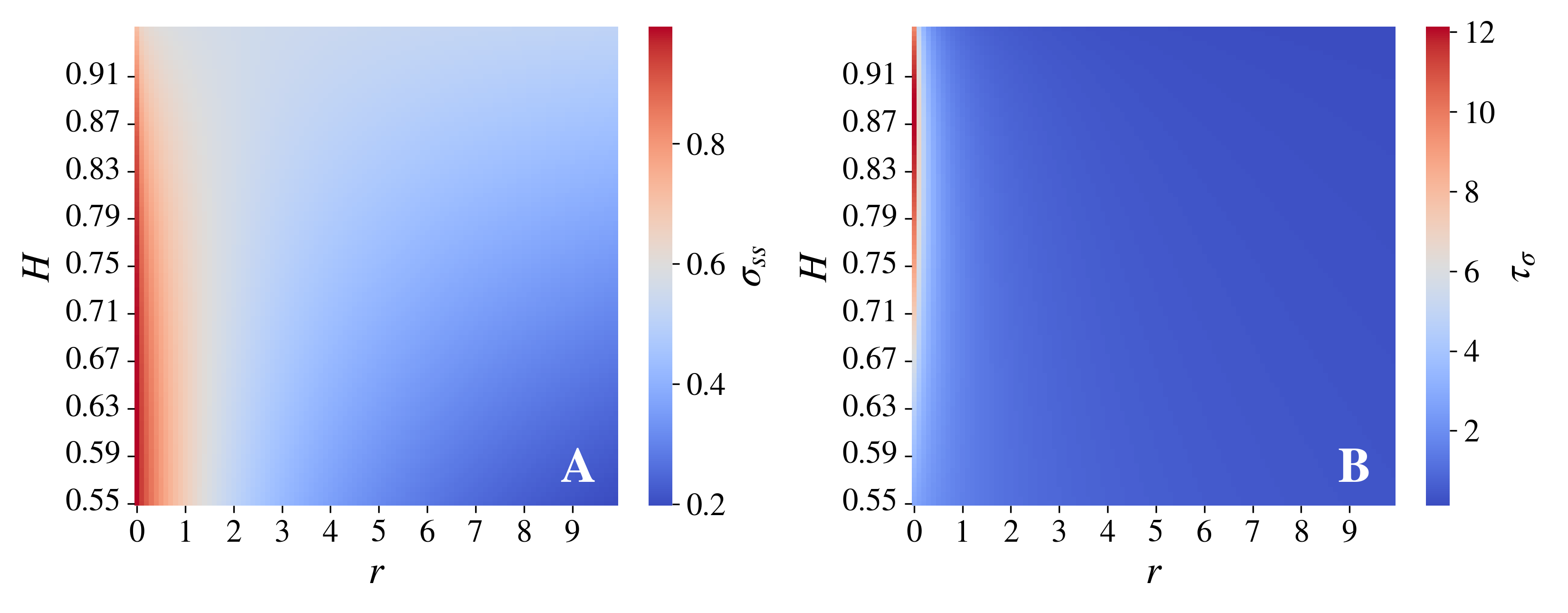}
    \caption{Effects of Hurst index and reset rate on (A) MSD saturation value ($\sigma_{ss}$), and (B) corresponding saturation timescale ($\tau_{\sigma}$) of the underlying process.}
    \label{fig:msd_saturation}
\end{figure}

\noindent
The dynamics are saturated under resetting, and the saturation timescale depends on the memory kernel, as well as on the reset rate. A recent study showed that the MSD of particles in a viscoelastic bath has an intermediate plateau, which was a result of the viscoelastic response as defined by the Jeffreys fluid model \cite{biswas2025}. However, upon applying resetting, the plateau disappeared \cite{biswas2025}. In the case of the power-law memory, we do not see any intermediate plateau. Instead, only two regimes exist: subdiffusion and saturation (due to the potential trap). We now explore these regimes as well as the corresponding effects when resetting is applied. The MSD under stochastic resetting can be written using the renewal method as
\begin{equation}
\label{eq:msd_with_resetting}
    \sigma^{2}_{r}(t) = e^{-rt} \sigma^{2}(t) + r \int_{0}^{t} e^{-r\tau} \sigma^{2}(\tau) ~d\tau,
\end{equation}
where $\sigma^{2}_{r}(t)$ represents the MSD under resetting. Now, by using the known results of $\sigma^{2}(t)$ for the no-resetting case (Eqs. \ref{eq:msd_no_reset_omega_neq_zero} and \ref{eq:msd_no_reset_omega_eq_zero}) in Eq. \eqref{eq:msd_with_resetting}, we get the MSD of the process under resetting. Although obtaining an analytical expression of the MSD is non-trivial when resetting is considered, it is possible to do so for the limiting case of free diffusion ($\omega = 0$). If we use Eq. \eqref{eq:msd_no_reset_omega_eq_zero} in the renewal equation, we get
\begin{equation}
\label{eq:msd_with_reset_omega_eq_zero}
    \sigma^{2}_{r}(t)\Big|_{\omega=0} = 
    \begin{cases}
        \frac{2k_{\text{B}}T}{\zeta A_{1}\Gamma(b+1)} \left[ \textrm{e}^{-rt} t^{b} + \frac{\Gamma(b+1) - \gamma(b+1, rt)}{r^{b+1}} \right], & \text{for finite} ~t ,\\
        \frac{2k_{\text{B}}T}{\zeta A_{1} r^{b+1}}, & \text{for } t \to \infty, 
    \end{cases}
\end{equation}
where, $\gamma(n,x) = \int_{0}^{x} e^{-t} t^{n-1} dt$ is the incomplete gamma function. The corresponding resetting-induced stationary states are shown in Fig. \ref{fig:msd}\bl{B}, \ref{fig:msd}\bl{C}, and \ref{fig:msd}\bl{D}. As discussed earlier, the MSD of free diffusion grows over time on a log-log scale, with a constant slope that depends on the Hurst index. However, the dynamics reach a stationary state (SS) at long times when resetting is considered \cite{evans2011, evans2011diffusion-optimal}, which is shown in Fig. \ref{fig:msd}\bl{B}. As we increase the reset rate, the convergence to SS is faster, and the curve saturates at relatively lower MSD amplitudes. Interestingly, resetting introduces an additional timescale of order $\tau_{r} \sim 1/r$, as is evident from the figure as well. However, it should be noted that the saturation happens as a combined outcome of the reset rate and the Hurst index. The MSD curve starts saturating at times that are proportional to the inverse of the reset rate. Although the dynamics are saturated at long times, they follow linear characteristics at short intervals, showing that subdiffusive behaviours are prevalent at short time scales, whereas resetting dominates over longer times. The similar effect of MSD under a harmonic trap ($\omega = 1$) and under resetting is shown in Fig. \ref{fig:msd}\bl{C}. A steady state occurs at an MSD amplitude of $k_{\text{B}}T/m \omega^{2}$ due to the harmonic trap. Resetting helps the system reach the SS faster, and the amplitude depends on the reset rate, as in the case of without any trap. Similarly, the effects of resetting and memory on MSD are shown in Fig. \ref{fig:msd}\bl{D}, where the figures are plotted at a fixed reset rate ($r=1$) but with different Hurst indices. Higher $H$ leads to delayed saturation of MSD when no resetting is considered, as shown in Fig. \ref{fig:msd}\bl{A} (coloured plots). But when the dynamics are restarted at a fixed reset rate, MSDs at different $H$ appear to be saturated at approximately the same timescale. This shows that the saturation time defined by memory is suppressed by resetting, as temporal correlations disappear. 
\\~\\
To better understand the effects of memory and resetting on MSD saturation, we plot the saturated MSD and the corresponding time scales in Fig. \ref{fig:msd_saturation}. In Fig. \ref{fig:msd_saturation}\bl{A}, we show a heat map of saturated MSD ($\sigma_{ss}$) as a function of $H$ and $r$. In the absence of resetting, MSD saturates at a lower value for higher $H$. On the other hand, in the presence of resetting, MSD saturates at a lower value for a higher reset rate, when the Hurst index is kept fixed. Lowering of MSD represents smaller effective diffusivity, either due to stronger memory correlation or stochastic resetting. These arguments match well with the results presented in Fig. \ref{fig:msd}.  
\\~\\
Further, as discussed earlier, the process leads to a steady state under trapping and resetting (though the nature of the steady states differ). Therefore, it is of interest to investigate the timescales of saturation and the combined effects of memory and resetting. To estimate the timescale, we calculate both the saturated MSD ($\sigma_{ss}$) and time-dependent MSD ($\sigma(t)$) for any given $H$ and $r$, and measure the difference $\Delta \sigma(t) = \sigma_{ss} - \sigma(t)$ at time $t$. When the difference becomes less than 1\% relative to $\sigma_{ss}$, \emph{i.e.}, $\sigma(t) \geq 0.99 ~\sigma_{ss}$, we consider the corresponding timescale as the saturation timescale of the dynamics. This is a purely computational approach, which has been adopted to avoid complicated analytics. Also, the 99\% cut-off does not have any special significance; one can perform the same analysis with 90\% or 95\% as the threshold. The important observation is that the saturation happens at a delayed time interval for higher $H$, when no resetting is considered. But under resetting, the dynamics saturate rapidly for any fixed $H$ value. Therefore, the study reveals an important insight into the saturation of the dynamics with associated timescales, and the dynamic effects of memory and resetting. 
\\~\\
The analytical MSDs are compared with numerical results, where MSDs are calculated from the trajectories simulated using Eq. \eqref{eq:final-numerical-solution}. We verified convergence of the numerical scheme with different $\Delta t$ values. Smaller $\Delta t$ has better convergence, but it is computationally expensive. Overall, a choice of $\Delta t = 10^{-3}$ has good convergence with analytical results over the preferred set (or range) of parameter values used in this work. 

\subsection{Correlation function under resetting} 
\label{sec:correlation_function} 
\begin{figure}
    \centering
    \includegraphics[width=0.9\linewidth]{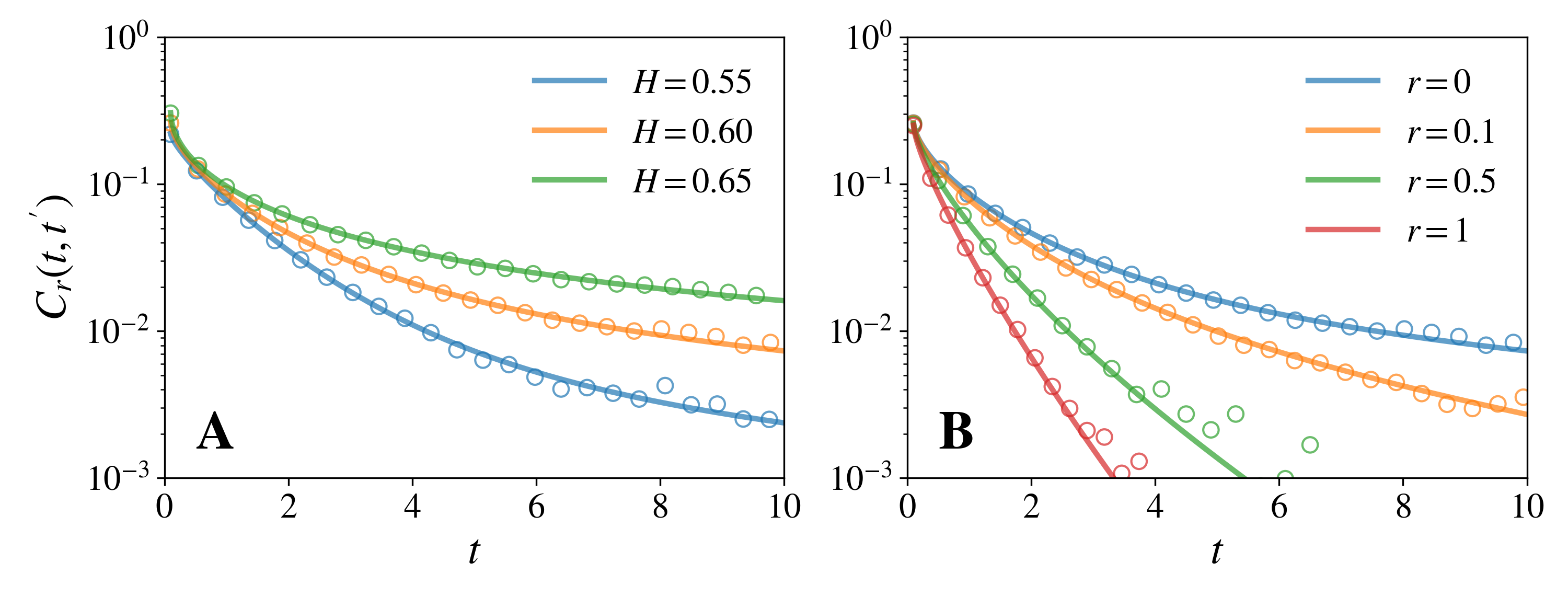}
    \caption{Correlation function $C_{r}(t,t')$ of a particle moving in a viscoelastic bath under resetting. (A) Correlation function of the actual dynamics at different Hurst indices without resetting ($r = 0$). (B) Correlation function of the dynamics under resetting with different reset rates. The Hurst index is fixed at $H = 0.6$. In both figures, we set $t' = 0.1$. The empty circles are estimated from $5 \times 10^{5}$ simulated trajectories with $\Delta t = 10^{-3}$.}
    \label{fig:correlation_function}
\end{figure}

\noindent
As we are working with correlated dynamics, it is important to know how the correlation function behaves under different correlation strengths, and the underlying impact when resetting is introduced \cite{vinod2022, biswas2025}. The position correlation function, as described in this work, is defined as $C(t,t') = \langle x(t) x(t') \rangle$, as mentioned in Eq. \eqref{eq:position_correlation_function}. Since the process is started from the origin ($x_{0}=0$), the first term vanishes, and the correlation function of the reset-free GLE becomes
\begin{equation}
\label{eq:correlation_function_without_resetting}
    C(t,t') = \frac{k_{\text{B}}T}{\zeta} \left[ I(t) + I(t') - I(|t-t'|) - \omega'^{2} I(t) I(t') \right].
\end{equation}
The corresponding figures are shown in Fig. \ref{fig:correlation_function}\bl{A}, which presents correlation functions (without resetting) for different Hurst indices. Higher $H$ corresponds to more strongly correlated dynamics and a larger correlation timescale, which is validated by the figure. 
\\~\\
Under resetting, the correlation function satisfies the renewal equation for $0 < t' < t$ and can be written as
\begin{equation}
\label{eq:correlation_function_with_resetting}
\begin{split}
    C_{r}(t,t') = e^{-rt} C(t,t') &\,+\, r \int_{0}^{t'} d\tau ~e^{-r(t-t'+\tau)} C(\tau,t-t'+\tau) \\ 
    &\,+\, r \int_{0}^{t-t'} d\tau ~e^{-r\tau} \langle x(t') \rangle_{r} \langle x(\tau) \rangle.
\end{split}
\end{equation}
The first term on the right represents those trajectories that do not undergo resetting during the observation period $[0,t]$. So, the probability of no reset $e^{-rt}$ is multiplied by the correlation function of the actual kinetics $C(t,t')$. For the other set of trajectories, which experience resetting, there are two possibilities. One of the possibilities is that the resetting occurs before $t'$, which is represented by the second term on the right-hand side of Eq. \eqref{eq:correlation_function_with_resetting}. Consider the last reset occurs $\tau$ time before $t'$, \emph{i.e.}, at time $t'-\tau$, where $0 < \tau < t'$, followed by diffusion upto time $t$. Thus, measuring from the last reset event, the two time instants $t'$ and $t$ correspond to the intervals $\tau$ and $t-t'+\tau$, and thus the underlying correlation function becomes $C(\tau,t-t'+\tau)$. The factor $r d\tau$ represents the probability of reset in the $d\tau$ interval, multiplied by the probability of no reset $e^{-r(t-t'+\tau)}$ for the rest of the interval $t-t'+\tau$, and integrated over all possible values of $\tau$ in $[0,t']$. The third contribution comes from the trajectories which undergo a last resetting in $[t',t]$. Suppose the last reset happens $\tau$ time before $t$ with $0< \tau < t-t'$. Consider the positions of the particle before and after the reset to be $x(t')$ and $x(t)$. Therefore, $x(t')$ and $x(t)$ can no longer be described by a two-time correlation $C(t,t') = \langle x(t) x(t') \rangle$ of the same continuous trajectory. Instead, the correlation can now be factorized as $C(t,t') = \langle x(t')\rangle_{r} \langle x(\tau) \rangle$. The factor $rd\tau$ represents the probability of reset in the $d\tau$ interval, followed by a no-reset probability $e^{-r\tau}$ for the remaining time $\tau$, and integrating over all possible $\tau$ values. For simplicity, we consider $x_{0} = x_{r} = 0$, which leads to the vanishing of the third term on the right-hand side and simplifies the expression to 
\begin{equation}
\label{eq:correlation_function_with_resetting_simplified}
    C_{r}(t,t') = e^{-rt} C(t,t') \,+\, r \int_{0}^{t'} d\tau ~e^{-r(t-t'+\tau)} C(\tau,t-t'+\tau) .
\end{equation}
The corresponding results are shown in Fig. \ref{fig:correlation_function}\bl{B}, in which the correlation functions for different values of the reset rate are shown, keeping $H$ fixed at 0.6. When resetting is absent, the dynamics has strong correlation with some characteristic correlation time. As we reset the position and bath memory, the intermittent correlation disappears. How quickly the disappearance happens depends on how frequently the dynamics is restarted, which is represented by different reset rates, as shown in Fig. \ref{fig:correlation_function}\bl{B}. This analysis has direct implications for the first-passage behaviours and the emergence of an exponential decay of the FPT distributions, characterized by the elimination of long-tailed rare-event trajectories due to an effective memory loss by resetting.

\subsection{Resetting of the memory kernel} 
\label{sec:resetting_memory_kernel} 
\noindent
The renewal process enforces a complete resetting of the dynamics, which involves resetting the position of the particle, as well as the memory kernel. Therefore, it is important to also consider how the memory is being affected when resetting is introduced. Although the effect has been reported previously on a truncated power-law kernel \cite{jolakoski2025}, here we mention this for a complete understanding of the dynamics, which will be useful later to understand the first-passage behaviours of the system, and how resetting and memory, as a combination, determine the escape kinetics. The memory kernel under resetting, $K_{r}(t)$, can be described by the renewal equation, 
\begin{equation}
\label{eq:memory-kernel-under-resetting}
    K_{r}(t) = e^{-rt} K(t) + r\int_{0}^{t} d\tau ~e^{-r\tau} ~K(\tau). 
\end{equation}
\begin{figure}
    \centering
    \includegraphics[width=\linewidth]{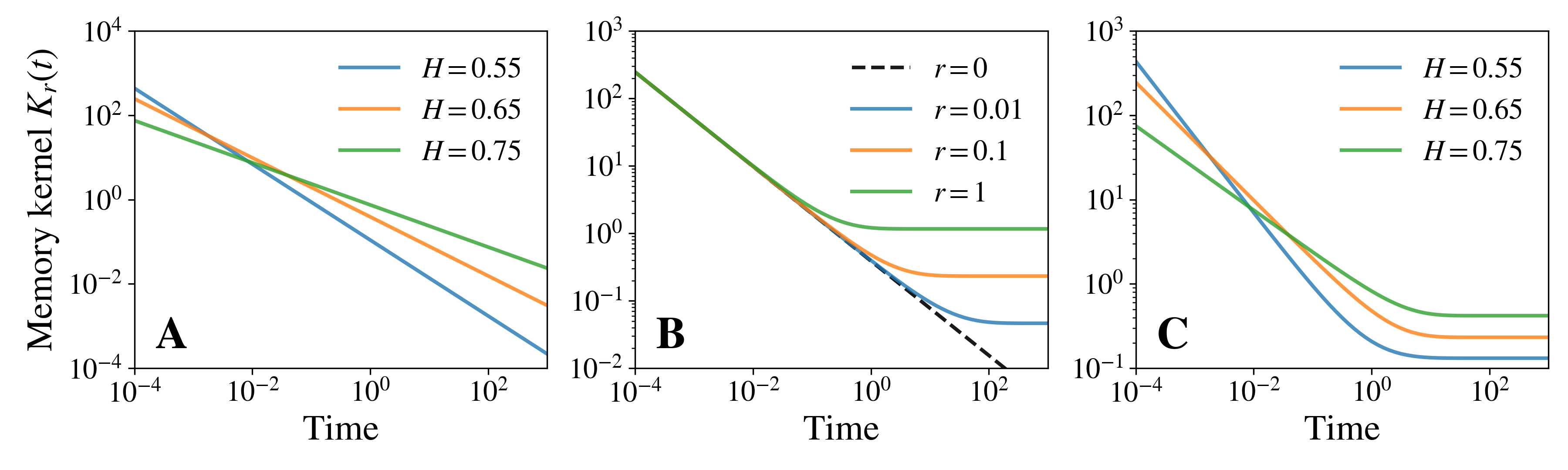}
    \caption{Effect of resetting on the power-law memory kernel. Kernel $K_{r}(t)$ at (A) different Hurst index values without resetting, (B) different reset rates at $H = 0.65$, and (C) fixed reset rate ($r = 0.1$) with different Hurst indices.}
    \label{fig:memory_kernel}
\end{figure} 
\\
The underlying effects are shown in Fig. \ref{fig:memory_kernel}, in which the memory kernel as a function of time is shown. In Fig. \ref{fig:memory_kernel}\bl{A}, we show how memory decays over time at different correlation strengths defined by the Hurst index ($H$). The memory decays faster when the noise is weakly correlated, whereas it decays slowly when the correlation is stronger. But at any time instant, a nonzero memory persists; no matter how large the time is, the system almost never relaxes. However, when the kernel is restarted (under complete renewal), the long-time correlation vanishes, and the memory gets saturated to a non-zero value. Interestingly, the saturation timescale is approximately equal to the inverse of the reset rate, \emph{i.e.}, $\sim 1/r$, as seen from Fig. \ref{fig:memory_kernel}\bl{B}. It should be noted that, with increasing reset rate, the saturation happens faster and at a higher value. Finally, in Fig. \ref{fig:memory_kernel}\bl{C}, we show how the memory kernel varies at different $H$, at a fixed reset rate. The kernel at a larger Hurst index decays more slowly and saturates at a relatively higher value. The time at which the kernel saturates depends on $H$, as the saturation appears at a later time for higher $H$. This property will be helpful to understand the first-passage behaviours of the particle escaping from the well, as well as the combined effects of memory and resetting.

\section{First-passage properties under resetting}
\label{sec:first-passage-properties}
\noindent
So far, we have discussed the dynamical properties of the system and its effects when resetting is considered. The probability as defined in Eqs. \eqref{eq:fpe-with-resetting} and \eqref{eq:renewal-equation} correspond to an unbounded system where a particle can diffuse in an infinite space and occasionally undergoes resetting. But when we are concerned about studying kinetic processes, we need to examine the statistics of particles escaping a stable potential well activated by thermal fluctuations. To investigate this, we evaluate the first-passage properties of the system considering an absorbing boundary placed away from the well minimum. Particles start diffusing from the well minimum and hit the boundary at random intervals of time. By repeating the process for a large sample of trajectories, we can eventually estimate the FPT distribution and the mean first-passage time (MFPT) of the process. It is known that the FPTs of many Markovian processes are exponentially distributed, as it corresponds to the escape of particles from the trap in the absence of correlation \cite{kramers1940, hanggi1986, hanggi1990}. However, in non-Markovian systems, the FPT distributions are characterized by heavy tails (due to the presence of rare-event trajectories, that take an unusually long time to escape), which typically reflect the correlated nature of the dynamics \cite{chaudhury2006, wiggin2008, vilk2024}. Therefore, it remains to be seen what effect stochastic resetting has on the first-passage behaviour of the system, and how the correlated long-tailed behaviours are modulated when the dynamics are restarted. 
\\~\\
In the presence of an absorbing boundary, the probabilities, as defined in Eqs. \eqref{eq:fpe} and \eqref{eq:renewal-equation}, are no longer conserved. As particles escape from the well, the total probability reduces over time. Suppose, $P_{a}(x,t|x_{0})$ is the probability at time $t$ in the confined space $\Omega \in (-\infty, x_{a}]$, such that $P_{a}(x\to -\infty, t|x_{0}) = 0$, and $P_{a}(x=x_{a}, t|x_{0}) = 0$, where $x_{a}$ is the position of the absorbing boundary. The survival probability at time $t$ can then be defined by integrating $P_{a}(x,t|x_{0})$ over the entire region $\Omega$, as
\begin{equation}
\label{eq:survival_prob}
    Q(t|x_{0}) = \int_{\Omega} P_{a}(x,t|x_{0}) dx.
\end{equation}
The survival probability, $Q(t|x_{0})$, then represents the particle's probability of not getting absorbed by the boundary at time $t$, starting from $x_{0}$. After finding the survival probability, we can evaluate the first-passage time (FPT) distribution, $f(t|x_{0})$, denoting the distribution of times required by the particles to escape from the well. This can be calculated as
\begin{equation}
\label{eq:waiting-time-distribution}
    f(t|x_{0}) = -\frac{d}{dt} Q(t|x_{0}).
\end{equation}
Once we obtain the FPT distribution, we can calculate the MFPT by evaluating the first moment of $f(t|x_{0})$, given as, 
\begin{equation}
\label{eq:mfpt}
    \big\langle T \big\rangle = \int_{0}^{\infty} t~ f(t|x_{0}) dt = -\int_{0}^{\infty} t \frac{d}{dt} Q(t|x_{0}) dt = \int_{0}^{\infty} Q(t|x_{0}) dt.
\end{equation}
using the conditions $Q(0|x_{0})=1$, and $Q(\infty|x_{0})=0$.

\subsection{First-passage properties under resetting} 
\begin{figure}
    \centering
    \includegraphics[width=\linewidth]{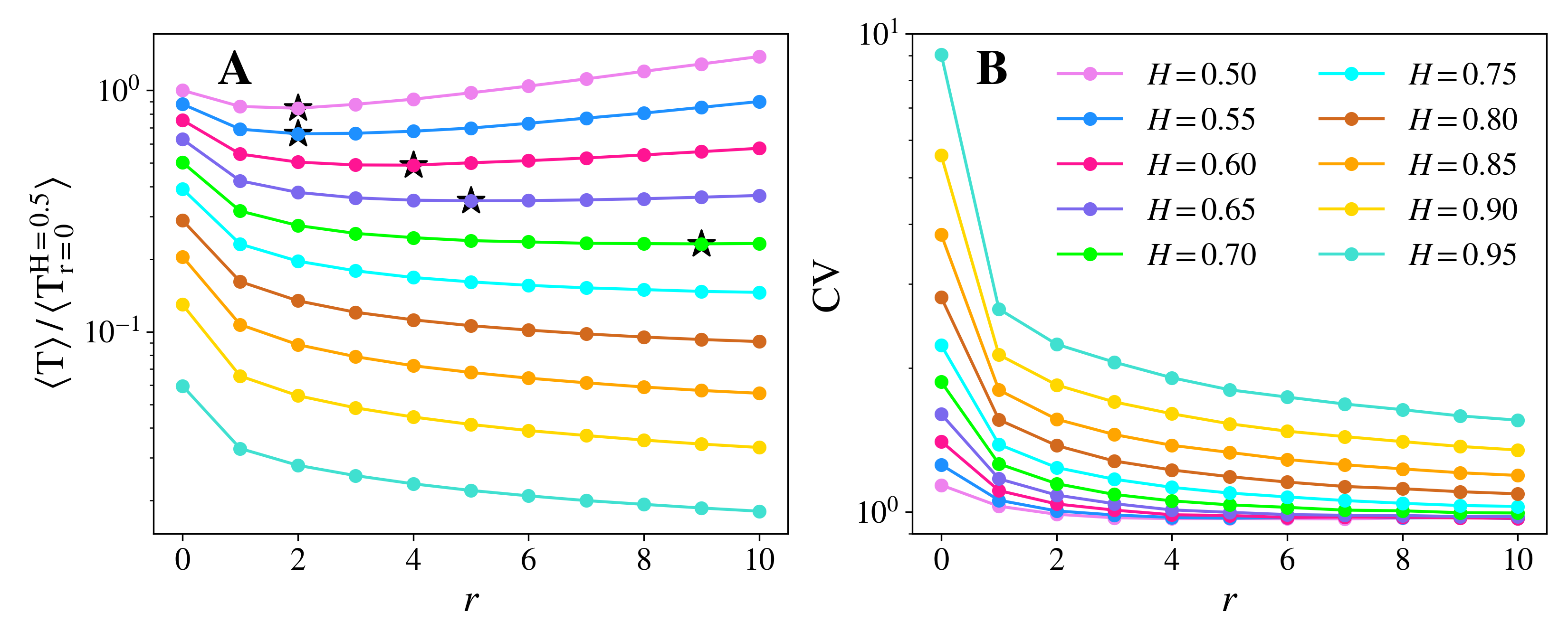}
    \caption{Mean first-passage time and coefficient of variance of the FPT distributions of escape kinetics under resetting. (A) Scaled MFPT, $\langle T_{r} \rangle / \langle T^{H=0.5}_{r=0} \rangle$, as a function of reset rate at different values of the Hurst index. $\langle T^{H=0.5}_{r=0} \rangle$ is the MFPT for $r=0$ at the Markovian limit. The stars represent the optimal reset conditions. (B) Coefficient of variance (CV) of the FPT distributions as a function of reset rate at different Hurst indices. 
    The colour scheme in Fig. A is the same as that of Fig. B. The absorbing boundary was fixed at $x_{a}=1$, and resetting was performed at the well minimum, \emph{i.e.}, $x_{r}=0$. Reported values were estimated by simulating $10^{5}$ trajectories with $\Delta t = 10^{-3}$.}
    \label{fig:mfpt}
\end{figure}

\noindent
The probability density as described by Eq. \eqref{eq:renewal-equation} is true for unbounded systems, in which we mostly focus on the dynamics of the system. But when we deal with the first-passage behaviours, the description needs to be modified. In the presence of an absorbing boundary, the renewal equation for the survival probability is expressed as
\begin{equation}
\label{eq:renewal-equation-with-absorbing-boundary}
    Q_{r}(t|x_{0}) = e^{-rt} Q_{0}(t|x_{0}) + r\int_{0}^{t} d\tau ~e^{-r\tau} Q_{0}(\tau | x_{r}) Q_{r}(t-\tau | x_{0}),
\end{equation}
where $Q_{r}(t|x_{0})$ represents the survival probability at time $t$ with resetting, and $Q_{0}(t|x_{0})$ is the survival probability without any resetting. The first term on the right corresponds to the trajectories that do not experience any reset for the entire time $t$, multiplied by the probability of no reset, $e^{-rt}$. The second term combines the trajectories that have experienced at least one reset. It is a convolution of two survival probabilities, the survival probability $Q_{r}$ up to time $t-\tau$ starting from $x_{0}$ and undergoing resetting, and the survival probability $Q_{0}$ starting from $x_{r}$ and diffusion without resetting for the rest of the $\tau$ interval. Taking the Laplace transform of Eq. \eqref{eq:renewal-equation-with-absorbing-boundary}, we get 
\begin{equation}
\label{eq:renewal_survival}
    \hat{Q}_{r}(s|x_{0}) = \frac{\hat{Q}_{0}(r+s|x_{0})}{1 - r \hat{Q}_{0}(r+s | x_{r})}
\end{equation}
where, $\hat{f}(s) = \int_{0}^{\infty} e^{-st} f(t) dt$ represents the Laplace transform of $f(t)$. 
\\~\\
To solve Eq. \eqref{eq:renewal_survival}, we first need to evaluate $Q_{0}(t|x_{0})$, the survival probability of the original process (without resetting), and then use it to get $Q_{r}(t|x_{0})$. But, solving $Q_{0}(t|x_{0})$ requires the propagator of the underlying process by explicitly considering the absorbing boundary, which is complex and non-trivial. To avoid this complication, we study the first-passage properties numerically, at different values of reset rates ($r$) and Hurst indices ($H$). This will lead to understanding the interplay between the memory effect (generating a persistent temporal correlation and causing delayed escape) and the resetting that is known to accelerate the dynamics. 
\\~\\
The numerical scheme is similar to that used to understand the system's dynamics, given by the discretized version of the GLE, as mentioned in Eq. \eqref{eq:final-numerical-solution}. To evaluate first-passage behaviours, we placed an absorbing boundary at $x = x_{a}$, and considered that as soon as the particle hits the boundary, it is immediately removed from the system. Numerically, it has been implemented as follows: if the position of the particle $x(t) \geq x_{a}$, the simulation is terminated, and the corresponding time $t$ is considered the first-passage time (FPT) of the trajectory. We simulated a large ensemble of trajectories (in the order of $10^{5}$ or $10^{6}$), with an appropriate choice of time interval ($\Delta t = 10^{-3}$), to obtain an approximate realization of the first-passage behaviours. To find the most suitable choice of $\Delta t$, we checked convergence of MFPT values at different $\Delta t$ values, and $\Delta t = 10^{-3}$ appears to be a reasonable choice. A smaller $\Delta t$ would give more accurate results, but that would be computationally more expensive. 
\\~\\
The Kramer's escape in a non-Markovian setup is characterised by non-exponential decay of FPT distributions with the presence of long tails. This indicates that although many of the trajectories escape the well within a very short time, some trajectories take much longer to escape, corresponding to the rare events \cite{chaudhury2006, chatterjee2010, kuo2010, chatterjee2011, barbier2024}. Recent studies have shown that, by introducing stochastic resetting to molecular dynamics simulations, structural transitions from one state to another can be accelerated by several orders of magnitude compared to the actual process \cite{blumer2022, blumer2024}. In such complex processes, particles may get trapped in some metastable states and take longer to escape, leading to broad transition time distributions (times required for a transition from one state to another) similar to the FPT distributions in our case. By resetting the dynamics, those trapped particles get another chance to start over and escape in a much shorter time, eventually making the distributions narrower \cite{blumer2022}. Ideally, it is expected that for an FPT distribution with a coefficient of variance ($\text{CV} = \sigma/\mu$, where $\sigma$ is the standard deviation, and $\mu$ is the mean of the distribution) greater than unity, resetting can be beneficial to accelerate the dynamics \cite{pal2022, blumer2024}. 
\begin{figure}
    \centering
    \includegraphics[width=\linewidth]{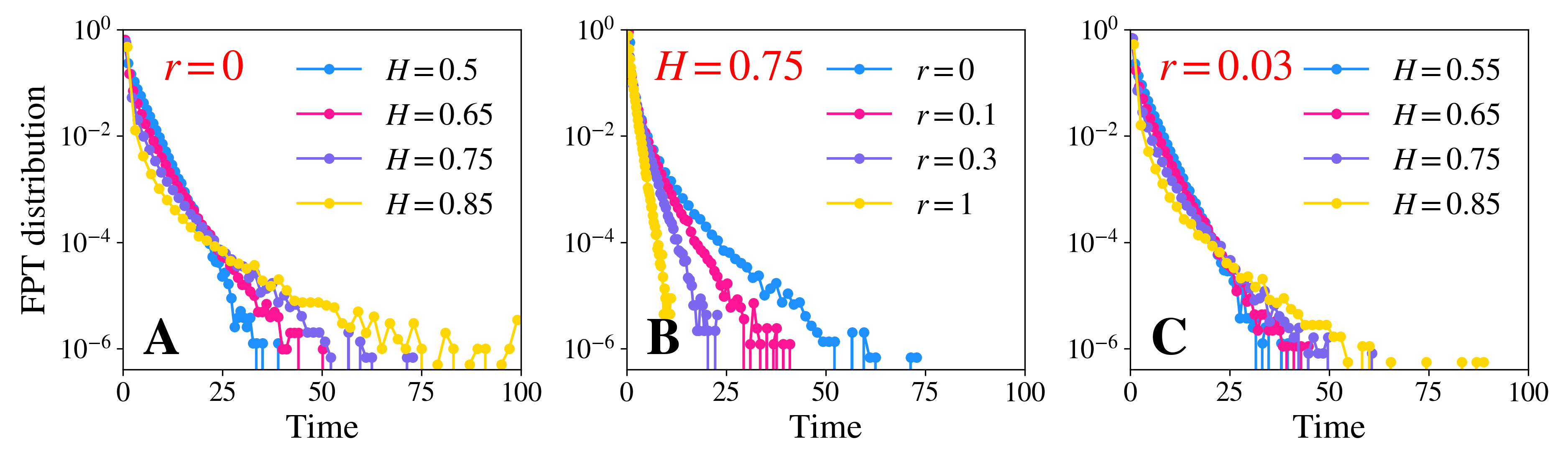}
    \caption{First-passage time distributions of particles escaping from a potential well in a viscoelastic bath. (A) FPT distributions of the escape kinetics without any resetting ($r = 0$), at different Hurst indices. (B) FPT distributions under Poissonian resetting with different reset rates at $H = 0.75$. (C) Distributions at a fixed reset rate ($r = 0.03$) and different Hurst indices. FPTs were estimated from $10^{6}$ trajectories with $\Delta t = 10^{-3}$. For all the figures, we fixed $x_{a}=1$, and resetting position at $x_{r} = 0$.}
    \label{fig:fpt-distributions} 
\end{figure}
\\~\\
The FPT distributions are shown in Fig. \ref{fig:fpt-distributions}\bl{A}, corresponding to different Hurst indices and no-reset ($r = 0$). They appear with long tails having $\text{CV} > 1$ (as shown in Fig. \ref{fig:mfpt}, which we will discuss later) because of the rare events. The distribution at the Markovian limit ($H = 0.5$, shown with the blue curve) shows exponential decay, as the particles escape from the well at random intervals, and no correlation is present. The higher the value of $H$, the broader the distribution becomes. 
The effects of resetting on the first-passage behaviours are presented in Fig. \ref{fig:fpt-distributions}\bl{B}. When resetting is applied, at a fixed $H$, the distribution becomes narrower, resulting in faster escape from the well, and effectively reducing the probabilities of rare events. With further increase in reset rates, the distributions keep getting narrower. Importantly, as the reset rate increases, the non-exponential nature, particularly at long times, is reduced. However, at short times the distribution remains non-exponential, as evident from the figure. This indicates that in the short-time limit, the dynamics remain correlated; however, at long times, the correlation breaks as resetting is implemented. The timescale till which the distribution is non-exponential depends on the reset rate and the Hurst index. But, for a given value of $H$, the crossover from non-exponential to exponential happens more quickly for higher reset rates. This can be understood intuitively. As we reset the dynamics at random intervals, the temporal correlation breaks, which helps the rare-event trajectories to escape at a faster rate, leading to a narrow and exponential distribution. But trajectories that escape the well at very short times (at a timescale less than $\sim 1/r$) have negligible probability of experiencing resetting, and hence the memory correlation persists. Next, we focus on understanding the effects of memory on the first-passage behaviours in the presence of resetting. In Fig. \ref{fig:fpt-distributions}\bl{C}, we show the FPT distribution at fixed reset rates ($r=0.03$) and different Hurst indices. Higher $H$ leads to longer tails in the distribution, similar to those shown in Fig. \ref{fig:fpt-distributions}\bl{A}. For weak correlation, the decay is approximately exponential, and it becomes more non-exponential as the correlation becomes stronger. 
%
\\~\\
In addition to studying FPT distributions, it is also important to consider how the distribution properties, such as the MFPT and CV, are affected due to resetting. We plot these quantities for increasing reset rate and at different Hurst indices, which are shown in Fig. \ref{fig:mfpt}. It can be clearly seen from Fig. \ref{fig:mfpt}\bl{A} that MFPT varies non-monotonically, as is also the case for many target search systems where optimal reset conditions emerge, leading to a minimum MFPT \cite{evans2011diffusion-optimal, evans2020}. The optimal reset rates corresponding to the individual curves are shown with black asterisks. However, the non-monotonic variation of MFPT and obtaining an optimal target search condition is possible for a low Hurst index, \emph{i.e.}, when the dynamics are weakly correlated. For stronger correlations, the optimal reset conditions are reached at higher reset rates, except beyond $H > 0.75$, where the MFPT decreases steadily within the given observation range. One may notice that, as $ H$ increases, MFPT at any fixed reset rate decreases. It is possible that as the dynamics become more correlated, the particles diffuse persistently in the same direction and reach the target faster \cite{ferrer2021}. 
\\~\\
The variations of CVs of the FPT distributions show even more interesting properties. Usually, it is expected that for high variance, particularly when CV is greater than 1, resetting can enhance target search processes \cite{pal2022, blumer2022, blumer2024}. The broad FPT distributions, as a result of the delayed escape, become narrower when resetting is introduced, therefore reducing the CV of the distribution. In a recent molecular dynamics simulation study of complex molecules, it has been observed that resetting can speed up barrier crossing processes, and CV varies non-monotonically with reset rates \cite{blumer2024}. In contrast to that, CV in our case converges close to unity at large reset rates, for any value of the Hurst index parameter, as shown in Fig. \ref{fig:mfpt}\bl{B}. This happens because, as we reset the dynamics, a transition occurs from a non-exponential to an exponential distribution (see the FPT distribution in Fig. \ref{fig:fpt-distributions}\bl{B}). Further, the transition time also shortens with increasing $r$, leading to faster transitions. This implies that under resetting and after the transition, the FPT distribution follows Poissonian statistics, rendering the mean to be approximately equal to the standard deviation, thus resulting in CV converging to unity at higher reset rates. A Poissonian distribution indicates that the escape mechanism happens at random intervals, and no correlation exists. When resetting is applied, the variance of the distribution follows the same non-monotonic pattern as the mean (although the graph is not explicitly shown), and both of these quantities change in such a way that the CVs eventually converge to unity. This is another strong piece of evidence of a non-exponential to exponential crossover, suggestive of a transition from correlated to uncorrelated dynamics at large times. Another notable feature is that, at small $H$, the CV without resetting is already close to unity, and after resetting, the dynamics become approximately Markovian, even with a very small reset rate. But for large $H$, the FPT distributions have relatively broader tails initially, and it requires the dynamics to be restarted more frequently for the CV to converge to unity. Despite the convergence being very slow at higher $H$, frequent resetting can indeed accelerate the dynamics, but that would be energetically inefficient \cite{tal2020, olsen2024}. Therefore, the calculations corresponding to the first-passage properties suggest that the escape from metastable states can be accelerated by implementing stochastic resetting, and can be optimized by tuning the memory and reset frequency. 

\section{Discussions}
\label{sec:discussions}
\noindent
In this work, we studied the escape kinetics of a particle trapped in a potential well and under the influence of fractional Gaussian noise. The particle is immersed in a viscoelastic bath, leading to a non-Markovian process. The dynamics are intermittently restarted from the well minimum at Poissonian intervals to investigate whether resetting can be beneficial for faster escape from the well. Recent studies under a Markovian setup have shown that Kramer's escape process can be optimized by suitable resetting protocols to facilitate faster escape from the well \cite{ahmad2019, cantisan2021}. However, properties change when we deal with non-Markovian systems, as the dynamics are influenced by the bath memories, and therefore, there exists a trade-off between memory correlation and resetting. This has also been experimentally investigated for a colloidal bead diffusing in a viscoelastic fluid, specifically focusing on the first-passage behaviours under resetting, from which several key roles of memory and bath relaxation have been elucidated \cite{ginot2026}. This setup also shows that the relaxation time during each reset is important for the complete erasure of memory and the start of a fresh diffusion process, uncorrelated with the previous processes. 
\\~\\
In our current study of escape kinetics as well, the possible regimes where resetting can be beneficial over memory are shown. Certain key outcomes have been observed. First, there is an effective loss in long-term memory correlation, which leads to faster escape of rare-event trajectories when resetting is applied. It causes narrowing of the first-passage time distributions by eliminating the long tails, and enables a crossover from a non-exponential long-tailed distribution to a nearly exponential distribution of first-passage times. However, this crossover depends on the timescales of the reset rate and the correlation strength. This is observed in Fig. \ref{fig:mfpt}\bl{B}, where the CV converges to unity as $r$ increases, but the convergence rate depends on the Hurst index value. Interestingly, when a similar system involving a barrier crossing process is studied under a Markovian setup, the CV is found to vary non-monotonically with reset rate \cite{blumer2022}. But for non-Markovian systems, it converges to unity. The correlation function shows rapid decay under stochastic resetting, and becomes approximately exponential for higher reset rates, as shown in Fig. \ref{fig:correlation_function} \cite{biswas2025}. This can be another indication of the dynamics becoming uncorrelated, leading to escape at random intervals and an exponential FPT distribution. Second, the MPFT of non-Markovian escape kinetics reduces as resetting is considered, and optimal reset conditions can be found based on the minimum MFPT. However, the optimal reset condition depends on the memory correlation and requires more frequent resetting as correlation becomes stronger. But that can be energetically inefficient, as frequent resets would lead to an increase in cost to perform the task \cite{tal2020, olsen2024}. Additionally, we assumed that, during resetting, the bath is relaxed instantaneously, and the memory of past events is removed in zero time. This is certainly not true. In a memory-driven system, the bath needs to be relaxed after each reset, and the higher the correlation strength, the longer the relaxation time \cite{ginot2026}. Therefore, there exists a trade-off between effectiveness and energetic cost, and this is the deciding factor for whether frequent resetting is beneficial or not, over the actual dynamics \cite{tal2020, olsen2024}. 
Also, finding an optimal escape mechanism is non-trivial. In Markovian systems, optimization by resetting is characterised by a single timescale determined by $r^{\ast}$ \cite{evans2011, evans2011diffusion-optimal, evans2020}. But in a non-Markovian system, this optimization strongly depends on the noise correlation, and optimal escape for strong correlation is achieved at a high reset rate. 
\\~\\
The crossover from the long-tailed non-exponential distribution to an exponential distribution can indicate a possible transition from non-Markovian to Markovian dynamics, as the escape processes in memoryless dynamics happen at random times, leading to exponential distributions. But, if we look at the asymptotic limit of the memory kernel at long times, as shown in Fig. \ref{fig:memory_kernel}, it saturates to a non-zero value, and never decays to zero. This suggests that the dynamics under resetting are driven by a constant kernel and remain in the non-Markovian regime. Therefore, by only looking at the FPT distributions and CVs, it is not guaranteed that resetting causes a transition from non-Markovian to Markovian dynamics. All we can say is, in the actual kinetics (without resetting), particles escape from the well in a correlated manner, leading to rare events depending on the value of $H$. By resetting the dynamics, those rare event trajectories can escape the well faster, and the correlated escape is now transformed into escaping at random intervals. As we consider Poissonian resetting, the temporal correlation breaks abruptly, and the long-time decay becomes approximately exponential, indicating escape at random intervals. But, for very strong correlations, the bath needs to be reset very frequently to converge to an exponential distribution, as observed in the coefficient of variation plot.

\section{Conclusions}
\label{sec:conclusions}
\noindent
We formulate a combined analytical and numerical approach to incorporate stochastic resetting in non-Markovian systems to understand how resetting influences the dynamics and first-passage behaviours of the system. The escape kinetics of non-Markovian systems are characterized by non-exponential first-passage time distributions with heavy tails, caused by the delayed escape of rare-event trajectories. By resetting the dynamics, those rare-event trajectories can be made to escape from the well faster, thereby reducing the occurrence of rare events and effectively narrowing the FPT distributions. A crossover from non-exponential to exponential behaviour of the FPT distribution is observed, symbolising a discontinuity in temporal correlation, as complete resetting is implemented where both the position and memory get restarted. The study also suggests the presence of optimal reset conditions; however, the optimality is highly sensitive to the strength of dynamical correlations. It suggests that the optimal target search strategy by stochastic resetting is not always guaranteed, as the dynamics are coupled with the bath memory and depend on the coupling strength. This work, therefore, not only fills a crucial gap in the literature by incorporating stochastic resetting in memory-driven dynamics, but is also highly relevant in real systems like rapid bond-breaking, polymer translocation, anomalous diffusion and transportation in crowded environments, among others.


\section{Data and code availability}
\noindent
The data that support the findings of this study are available within the article. Relevant codes used in this work are available on GitHub at: \url{https://github.com/saha-debasish/escape-kinetics-with-resetting}.

\appendix

\section{Numerical solution}
\label{sec:numerical-method}
\noindent
In this section, we formulate a numerical method to solve Eq. (\ref{eq:gle-overdamped}) and validate the effect of stochastic resetting over diffusion in a potential well in the non-Markovian limit.
\\~\\
The GLE describing the equation of motion of the particle is
\begin{equation}
\label{eq:numerical-gle}
    m\omega^{2}x(t) = -\zeta \int_{0}^{t} dt^{\prime} K(t-t^{\prime}) \dot{x}(t^{\prime}) + \theta(t)
\end{equation}
where $K(t-t^{\prime})$ is the power-law memory kernel defined as 
\begin{equation}
\label{eq:numerical-memory-kernel}
    K(t-t^{\prime}) = 2H(2H-1) |t-t^{\prime}|^{2H-2}
\end{equation}
Using Eq. (\ref{eq:numerical-memory-kernel}) in Eq. (\ref{eq:numerical-gle}), we get
\begin{equation}
\label{eq:numerical-gle-2}
    m\omega^{2}x(t) = -\frac{A_{2}}{\Gamma(1-\alpha)} \int_{0}^{t} dt^{\prime} \frac{\dot{x}(t^{\prime})}{(t-t^{\prime})^{\alpha}} + \theta(t)
\end{equation}
where, $A_{2} = 2H(2H-1) \Gamma(1-\alpha) \zeta = A_{1} \zeta$ from Eq. \eqref{eq:relaxation_function}, and $\alpha = 2-2H$.
\\~\\
Now we use the Caputo derivative \cite{gorenflo1997}, which is defined for order $\alpha$ as 
\begin{equation}
\label{eq:caputo-derivative}
    ^{C}\mathcal{D}^{\alpha}_{0} \big(x(t)\big) = \frac{1}{\Gamma(1-\alpha)} \int_{0}^{t} dt^{\prime} \frac{\dot{x}(t^{\prime})}{(t-t^{\prime})^{\alpha}}
\end{equation}
Using this definition of Caputo derivative defined in Eq. (\ref{eq:caputo-derivative}) in Eq. (\ref{eq:numerical-gle-2}), we get
\begin{equation}
\label{eq:gle-to-caputo}
    ^{C}\mathcal{D}^{\alpha}_{0} \big(x(t)\big) = \frac{1}{A_{2}} \bigg(-m\omega^{2}x(t) + \sigma \dot{B}_{H} \bigg),
\end{equation}
where, $\dot{B}_{H}$ is the fractional Brownian noise with Hurst index $H$, and $\sigma$ is the noise strength defined as $\sigma = \sqrt{2k_{\text{B}}T\zeta}$ \cite{kou2004}. The fractional integral of order $\alpha$ is defined as \cite{gorenflo1997}
\begin{equation}
\label{eq:fractional-integral}
    ^{C}\mathcal{J}^{\alpha}_{0} \big(x(t)\big) = \frac{1}{\Gamma(\alpha)} \int_{0}^{t} dt^{\prime} (t-t^{\prime})^{\alpha - 1} x(t^{\prime}) 
\end{equation}
Integrating Eq. (\ref{eq:gle-to-caputo}) with the Caputo integral defined in Eq. (\ref{eq:fractional-integral}) and making use of the fundamental theorems of fractional calculus \cite{luchko2020, mainardi2007}
\begin{subequations}
    \begin{equation}
        ^{C}\mathcal{D}^{\alpha}_{0} \big(x(t)\big) = ~^{C}\mathcal{J}^{1-\alpha}_{0} \frac{dx}{dt}
    \end{equation}
    \begin{equation}
        ^{C}\mathcal{J}^{\alpha}_{0} ~^{C}\mathcal{D}^{\alpha}_{0} ~x(t) = x(t) - x(0)
    \end{equation}
\end{subequations}
we get,
\begin{equation}
\label{eq:solution-fractional-calculus}
    x(t) = x(0) + \frac{1}{A_{2}} \bigg[-\frac{m\omega^{2}} {\Gamma(\alpha)} \int_{0}^{t} dt^{\prime} (t-t^{\prime})^{\alpha-1} x(t^{\prime}) + \frac{\sigma}{\Gamma(\alpha)} \int_{0}^{t} (t-t^{\prime})^{\alpha-1} dB_{H} \bigg]
\end{equation}
We consider
\begin{equation}
\label{eq:fbm}
    G(t) = \frac{\sigma}{\Gamma(\alpha)} \int_{0}^{t} (t-t^{\prime})^{\alpha-1} dB_{H}
\end{equation}
where $G(t)$ is a Gaussian process of zero mean, as described in \cite{li2017, fang2020, batra2022}. Following \cite{li2017}, it can be shown that for $\alpha = 2-2H$ and $\sigma = \sqrt{2k_{\text{B}}T\zeta}$, the process $G(t)$ can be reduced to $G(t) \overset{d}{=} \beta_{H}B_{1-H}(t)$, where $\beta_{H} = \sigma \sqrt{\Gamma(3-\alpha)/\Gamma(1+\alpha)}$ and $\overset{d}{=}$ represents that both the processes have the same distribution. Therefore, $G(t)$ is a fractional Brownian motion process with Hurst index $1-H$. 
\\~\\
Now, we discretize Eq. (\ref{eq:solution-fractional-calculus}) by splitting the total time $t$ into $N$ equal segments with segment width of $\Delta t = t/N$. Approximating the integral in Eq. (\ref{eq:solution-fractional-calculus}) in the range between $(j-1)\Delta t$ and $j\Delta t$, and considering the function $x(t)$ as $x_{j-1}$, we get
\begin{equation}
    x_{N} = x(0) + \frac{1}{A_{2}} \bigg[ -\frac{m\omega^{2}}{\Gamma(\alpha)} \sum_{j=1}^{N} x_{j-1} \int_{(j-1)\Delta t}^{j\Delta t} \big(N\Delta t - t^{\prime}\big)^{\alpha-1} dt^{\prime} + G(N\Delta t) \bigg]
\end{equation}
Solving the integral, we finally get 
\begin{equation}
\label{eq:final-numerical-solution}
    x_{N} = x(0) + \frac{1}{A_{2}} \bigg[ -\frac{m\omega^{2}(\Delta t)^{\alpha}}{\Gamma(\alpha+1)} \sum_{j=1}^{N} x_{j-1} \Big\{ \big(N-j+1\big)^{\alpha} - \big(N-j\big)^{\alpha} \Big\} + G(N\Delta t) \bigg]
\end{equation}
This is the final equation used to simulate the trajectories and related properties like position distribution, mean-squared displacement, correlation function, and the first-passage properties.

\subsection{Validating $G(t) \overset{d}{=} \beta_{H}B_{1-H}(t)$}
\noindent
Now, let us check whether $G(t)$ can actually be represented by a fractional Brownian motion of index $1-H$. Following \cite{li2017}, it can be shown that $G(t)$ has the covariance
\begin{equation}
\label{eq:G_covariance}
    \left\langle G(t) G(t') \right\rangle = \frac{\beta_{H}^{2}}{2} \left[ t^{2(1-H)} + t'^{2(1-H)} - |t-t'|^{2(1-H)} \right].
\end{equation}
For $t = t'$, we get $\langle G(t)^{2} \rangle = \beta_{H}^{2} ~t^{2(1-H)}$. Next, we verify the stationary increment property, which is expressed as
\begin{equation}
    V(t,\tau) = \left\langle [G(t+\tau) - G(t)]^{2} \right\rangle.
\end{equation}
Making use of Eq. \eqref{eq:G_covariance}, and considering the initial condition that the process starts from the origin at $t=0$ simplifies the problem to 
\begin{equation}
\label{eq:stationary_increment}
    V(\tau) = \langle G(\tau)^{2} \rangle = \beta_{H}^{2} ~\tau^{2(1-H)},
\end{equation}
in which the increment $V(\tau)$ becomes stationary, \emph{i.e.}, independent of the time and depends only upon the interval $\tau$ for a fixed $H$. If we take $H=0.65$, then $\langle G(t)^{2} \rangle$ will be scaled as $t^{0.7}$, and $V(\tau)$ must be a constant depending on the choice of $\tau$. 
\begin{figure}
    \centering
    \includegraphics[width=\linewidth]{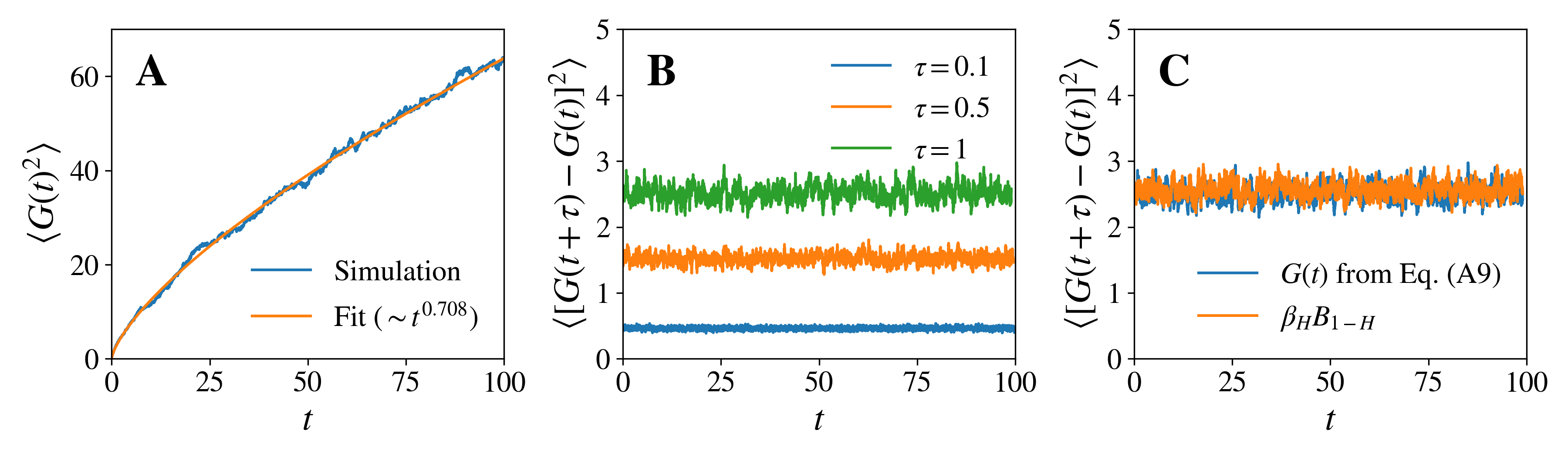}
    \caption{Evidence of $G(t)$, as defined in Eq. \eqref{eq:fbm}, is fractional Brownian motion with Hurst coefficient $1-H$. (A) Scaling of the covariance of $G(t)$ for $t=t'$. (B) Stationary increments of $G(t)$ at different $\tau$ values. (C) Comparison between $G(t)$ obtained from Eq. \eqref{eq:fbm} and $\beta_{H}B_{1-H}(t)$.} 
    \label{fig:fbm}
\end{figure}
\\~\\
To verify this, we generate the process $G(t)$ by numerically integrating Eq. \eqref{eq:fbm}, which consists of fractional Brownian motion of index $H$, and compare the scaling and stationary increment properties with respect to Eqs. \eqref{eq:G_covariance} and \eqref{eq:stationary_increment}, which are obtained from a fractional Brownian motion process with index $1-H$. The results are shown in Fig. \ref{fig:fbm}, which are generated at $H=0.65$. Fig. \ref{fig:fbm}\bl{A} shows how the covariance of $G(t)$ scales with time when $t = t'$. The blue curve is the trend that $\langle G(t)^{2} \rangle$ follows, which is simulated directly from Eq. \eqref{eq:fbm}. The fit (orange line) has an exponent of 0.708, which is approximately the same as the theoretical scaling value we calculated earlier from Eq. \eqref{eq:G_covariance}. 
\\~\\
Next, we show the stationary increment property of the process $G(t)$ in Fig. \ref{fig:fbm}\bl{B}. The stationary increment, as defined in Eq. \eqref{eq:stationary_increment}, is constant and only depends on $\tau$. In fact, one can calculate the values of $V(\tau)$ from Eq. \eqref{eq:stationary_increment} corresponding to different $\tau$ values. For example, if we take $\tau$ to be 0.1, 0.5 and 1, the corresponding saturation values will be 0.51, 1.58 and 2.57, respectively, which match well with the simulated results, as shown in the figure. Finally, we have shown a comparison of $G(t)$ from Eq. \eqref{eq:fbm} and $\beta_{H}B_{1-H}(t)$, which is generated from a fractional Brownian motion of index $1-H$, multiplied by the factor $\beta_{H}$. Both processes show a good match with each other, ensuring that both processes have similar properties.

\subsection{Markov limit}
\noindent
The above equation is correct, but diverges at the Markovian limit ($H = 1/2$) because of the parameter $A_{2}$. Therefore, we write a separate equation, discretize it, and calculate the FPT statistics for uncorrelated fluctuations. In the Markovian limit, memory kernel becomes uncorrelated, and the GLE in Eq. (\ref{eq:gle}) is reduced to to ordinary Langevin equation
\begin{equation}
    m \ddot{x}(t) = -\zeta \dot{x}(t) - m\omega^{2} x(t) + \theta(t)
\end{equation}
where $\theta(t)$ is the random force with mean zero, and variance $\left\langle \theta(t) \theta(t^{\prime}) \right\rangle = 2\zeta k_{\text{B}}T \delta(t - t^{\prime})$. In the overdamped limit, we can write
\begin{equation}
    \zeta \dot{x}(t) = - m\omega^{2} x(t) + \theta(t)
\end{equation}
which can be discretised using the Euler-Maruyama scheme \cite{maruyama1955, kloeden1977} as
\begin{equation}
    x(t + \Delta t) = x(t) - m\omega^{2}x(t) \Delta t + \sqrt{2\zeta k_{\text{B}}T \Delta t} ~\theta(t)
\end{equation}
We have used this equation to estimate FPT properties in the Markovian limit, whose results are shown in Figs. \ref{fig:fpt-distributions}\bl{A} and \ref{fig:mfpt}\bl{A}. 

\section{Simulation steps}
\label{app:simulation-steps}
\noindent
In section \ref{sec:numerical-method}, we discussed the method that we used to discretize the generalized Langevin equation (Eq. (\ref{eq:gle-overdamped})) using the Caputo derivative, which reduced to a discretized form mentioned in Eq. (\ref{eq:final-numerical-solution}). This equation is used to simulate trajectories of particles influenced by correlated fluctuations and the potential well. The simulation techniques are described as follows: 
\begin{enumerate}
    \item [1.] We created a two-dimensional NumPy array of dimension $ntrajs\times N$ to store the positions of the particles, where $ntrajs$ is the number of trajectories, and $N$ is the number of time steps for each trajectory. Therefore, it represents a 2D matrix with each element corresponding to the position of individual particles at different times. All the trajectories were started from the origin at $t = 0$.
    
    \item [2.] Trajectories are simulated in a time interval $t \in [t_{\text{init}}, t_{\text{fin}}]$. The simulations were run for $N$ steps, with a time interval $\Delta t = (t_{\text{fin}}-t_{\text{init}})/N$. We used $t_{\text{init}} = 0$ in every simulation.
    
    \item [3.] We generated Gaussian random variables following Eq. (\ref{eq:fbm}) with Hurst index of $1-H$. Different seeds were used for different trajectories. The $\tt{FractionalBrownianMotion}$ library from the module $\tt{stochastic}$ is used to generate the random variables \cite{stochastic2018}. Corresponding to each trajectory, we generated an array of random numbers having $N$ elements at the beginning, for the given time interval $[t_{\text{init}}, t_{\text{fin}}]$.
    
    \item [4.] Positions of the particles are updated using Eq. (\ref{eq:final-numerical-solution}).
    
    \item [5.] To introduce resetting, we call a random number $z \in [0,1]$ from the uniform distribution. At any time instant $t_{i}$, if $z < r\cdot dt$, the particle is brought back to the reset position $x_{r}$. Here, $r$ is the reset rate. A completely new diffusion process starts from the reset position at time $t_{i+1}$. We enforce that the memory of the previous process is also reset following the complete renewal protocol.
    
    \item [6.] As the reset happens at $i^{\text{th}}$ step, corresponding to a time $t_{\text{reset}}$, a new sequence of random numbers is generated following step 3 for the next diffusive phase. The new random numbers have the length of $N-i$ and for the time interval $[t_{\text{init}}+t_{\text{reset}}, t_{\text{fin}}]$. These random numbers are completely uncorrelated with the previous ones. Here, $t_{i}$ and $t_{\text{reset}}$ are used equivalently.
\end{enumerate}
These are the steps that are used to simulate the trajectories of particles diffusing in a potential well under correlated fluctuations. To obtain the FPT statistics, we put an absorbing boundary at $x=x_{a}$. As soon as a particle hits the absorbing boundary, the process is terminated, and the corresponding time is considered the FPT of the trajectory.

\section{Mean-squared displacement in viscoelastic bath}
\label{app:msd_derivation}
\noindent
The mean-squared displacement (MSD) can be calculated following the methods described in \cite{desposito2009, biswas2025}. We first take the Laplace transform of Eq. \eqref{eq:gle-overdamped} with the initial condition $x(0) = x_{0}$. The dynamic evolution in Laplace space after a few rearrangements can be expressed as 
%
\begin{equation}
    \hat{x}(s) = \frac{\hat{K}(s)}{s \hat{K}(s) + \omega'^{2}} x_{0} + \frac{1}{\zeta}\frac{\hat{\theta}(s)}{s \hat{K}(s) + \omega'^{2}},
\end{equation}
where $\omega'^{2} = m\omega^{2}/\zeta$. This equation can be modified as
\begin{equation}
\label{eq:eom_in_laplace_space}
    \hat{x}(s) = \left( \frac{1}{s} - \omega'^{2} \hat{I}(s) \right) x_{0} + \frac{1}{\zeta} \hat{G}(s) \hat{\theta}(s).
\end{equation}
Here, $G(t)$ is the relaxation functions $I(t)$ and $G(t)$ are the Laplace transforms of
\begin{equation}
\label{eq:relaxation_functions_G_and_I}
    \begin{split}
        \hat{G}(s) &= \frac{1}{s \hat{K}(s) + \omega'^{2}}, \\
        \hat{I}(s) &= \frac{s^{-1}}{s \hat{K}(s) + \omega'^{2}} = \frac{\hat{G}(s)}{s}.
    \end{split}
\end{equation}
Taking inverse Laplace transform of Eq. \eqref{eq:eom_in_laplace_space}, we get
\begin{equation}
    x(t) = \langle x(t) \rangle + \frac{1}{\zeta} \int_{0}^{t} dt' G(t-t') \theta(t'),
\end{equation}
where, 
\begin{equation}
\label{eq:position}
\begin{split}
    \langle x(t) \rangle = \mathcal{L}^{-1}\left[ \frac{1}{s} - \omega'^{2} \hat{I}(s) \right] x_{0} 
    = \left[ 1 - \omega'^{2} I(t) \right] x_{0}
\end{split}
\end{equation}
\\
Now making use of the double Laplace transform technique, Eq. \eqref{eq:eom_in_laplace_space} can be modified as
\begin{equation}
\label{eq:correlation_function}
    \left\langle \hat{x}(s) \hat{x}(s') \right\rangle = x_{0}^{2} ~\hat{\chi}(s)\hat{\chi}(s') + \frac{1}{\zeta^{2}} \hat{G}(s) \hat{G}(s') \left\langle \hat{\theta}(s) \hat{\theta}(s') \right\rangle,
\end{equation}
where we consider 
\begin{equation}
    \hat{\chi}(s) = \left[ \frac{1}{s} - \omega'^{2} \hat{I}(s)\right].
\end{equation}
\\
Following \cite{pottier2003}, it can be shown that 
\begin{equation}
    \left\langle \hat{\theta}(s) \hat{\theta}(s') \right\rangle = \zeta k_{\text{B}}T ~ \frac{\hat{K}(s) + \hat{K}(s')}{s+s'}
\end{equation}
which can be further simplified by replacing $\hat{K}(s)$ by $\hat{G}(s)$ from Eq. \eqref{eq:relaxation_functions_G_and_I}, which leads to
\begin{equation}
    \left\langle \hat{\theta}(s) \hat{\theta}(s') \right\rangle = \frac{\zeta k_{\text{B}}T}{s+s'} \left[ \frac{1}{s \hat{G}(s)} + \frac{1}{s' \hat{G}(s')} - \omega'^{2} \frac{s+s'}{ss'} \right]
\end{equation}
\\
Now, using the relation $\hat{G}(s) = s \hat{I}(s)$ in the above equation, and putting it back to Eq. \eqref{eq:correlation_function}, it becomes (after some rearrangements)
\begin{equation}
    \left\langle \hat{x}(s) \hat{x}(s') \right\rangle = x_{0}^{2} ~ \hat{\chi}(s)\hat{\chi}(s') + \frac{k_{\text{B}}T}{\zeta} \left[ \frac{\hat{I}(s)}{s'} + \frac{\hat{I}(s')}{s} - \frac{\hat{I}(s)+\hat{I}(s')}{s+s'} - \omega'^{2} \hat{I}(s) \hat{I}(s') \right]
\end{equation}
\\
The double inverse Laplace transform of the above equation leads to
\begin{equation}
    \left\langle x(t) x(t') \right\rangle = x_{0}^{2} ~ \chi(t)\chi(t') + \frac{k_{\text{B}}T}{\zeta} \left[ I(t) + I(t') - I(|t-t'|) - \omega'^{2} I(t) I(t') \right]
\end{equation}
which can be further written by replacing $\chi(t)$ as
\begin{equation}
\label{eq:position_correlation_function}
    \left\langle x(t) x(t') \right\rangle = \langle x(t) \rangle \langle x(t') \rangle + \frac{k_{\text{B}}T}{\zeta} \left[ I(t) + I(t') - I(|t-t'|) - \omega'^{2} I(t) I(t') \right]
\end{equation}
\\
The mean-squared displacement is represented as
\begin{equation}
\label{eq:msd_equation}
    \sigma^{2}(t) = \left\langle x^{2}(t) \right\rangle - \left\langle x(t) \right\rangle ^{2}
\end{equation} 
Using $t = t'$ in Eq. \eqref{eq:position_correlation_function}, and utilizing Eq. \eqref{eq:msd_equation}, we get
\begin{equation}
    \sigma^{2}(t) = \frac{k_{\text{B}}T}{\zeta} \left[ 2I(t) - \omega'^{2} I^{2}(t) \right]
\end{equation}
This is the final expression of the mean-squared displacement of a particle moving in a viscoelastic bath. This equation has been used to study the memory effect over the dynamics of the system and how resetting affects it in short and long time regimes, which are described in Sec. \ref{sec:msd}.

\bibliographystyle{apsrev4-1}
\bibliography{references}

\end{document}